\documentclass[review=false,
  screen=true,
natbib=false,
]{acmart}

\ifdefined\ifdebug \PassOptionsToPackage{showframe}{geometry}
  \ifLuaTeX \usepackage[All]{lua-typo}
  \fi \fi 

\usepackage{enumitem}
\usepackage{listings}
\lstdefinestyle{mystyle}{breaklines=true,
  basicstyle=\ttfamily\scriptsize,
  keepspaces=false,
}
\usepackage{custom-defs}
\usepackage{subcaption}

\usepackage[datamodel=acmdatamodel,
  style=acmnumeric,
  backend=biber,
]{biblatex}
\DeclareMathOperator{\embed}{embed}
\DeclareMathOperator{\iqr}{IQR}
\DeclareMathOperator{\md}{MD}

\NewDocumentCommand{\numSpritesPerLanguage}{}{\exnum{5000}\xspace}

\NewDocumentCommand{\numProjectsSpriteNaming}{}{\exnum{656164}\xspace}
\NewDocumentCommand{\numSpritesSpriteNamingEval}{}{\exnum{167713}\xspace}
\NewDocumentCommand{\numSpritesSpriteNamingTest}{}{\exnum{169187}\xspace}
\NewDocumentCommand{\numSpritesSpriteNamingTrain}{}{\exnum{1824786}\xspace}

\begin{document}

\title{EmbeddedKittens: An Evaluation of Code Embeddings for Scratch}

\author{Benedikt Fein}
\orcid{0000-0002-3798-845X}
\email{benedikt.fein@uni-passau.de}
\affiliation{\institution{University of Passau}
  \city{Passau}
  \country{Germany}
}
\author{Gordon Fraser}
\orcid{0000-0002-4364-6595}
\email{gordon.fraser@uni-passau.de}
\affiliation{\institution{University of Passau}
  \city{Passau}
  \country{Germany}
}

\setcopyright{cc}
\setcctype{by}
\acmJournal{TOSEM}

\begin{abstract}

The trend of embedding source code for machine learning applications
also enables new opportunities in learning analytics in programming
education, but which code embedding approach is most suitable for
learning analytics remains an open question.
A common approach to embedding source code lies in treating the code
as a token sequence similar to natural language when training large
language models~(\llm{}s). However, in case of visual block-based
programming languages like \Scratch, this approach cannot be applied
directly. While text-based representations of block-based code can be
created to apply \llm{}s to this problem, other dedicated embedding
models could potentially exhibit improved performance by capturing
additional structural information.
In this paper, we therefore instantiate four \llm{}s and five
different popular embedding approaches for \Scratch programs, create a
token-prediction and two different classification tasks with
corresponding datasets, and empirically evaluate the models on them.
Our experiments demonstrate that a transfer of code embeddings to the
educational environment of \Scratch is feasible. The embedding models
trained on large open \Scratch datasets capture relevant structural
and semantic information about the code to enable learning analytics
like predicting functional correctness of student programs, in the
typically small classroom setting without requiring further
task-specific model fine-tuning.

 \end{abstract}

\keywords{Scratch, code embeddings, programming education.}

\ccsdesc[300]{Social and professional topics~K-12 education}
\ccsdesc[500]{Social and professional topics~Software engineering education}
\ccsdesc[500]{Software and its engineering~Visual languages}
\ccsdesc[300]{Computing methodologies~Machine learning}

\maketitle

\section{Introduction}

The proliferation of code embeddings has led to their adoption for a
wide variety of software engineering tasks, especially since large
language models~(\llm{}s) have become widely
available~\cite{Hou2024Large}. Combining them with integrations into
coding agents~(\eg\ \properNoun{GitHub Copilot}, \properNoun{Claude
Code}) even allows developers to delegate tasks to these agents for
semi-autonomous resolution. The \llm{}s are trained on large datasets
of openly available natural language and source
code~\cite{Singh2026OpenAI}. Since open-source code datasets consist
primarily of commonly used text-based programming languages like
Python or Java~\cite{Kocetkov2022Stack,Lozhkov2024StarCoder}, the
\llm{}s are performing best on such common
languages~\cite{Joel2025Survey}. Thus, the main target audience of
such \llm-based tools are professional, hobbyist, or learning
developers using these textual programming languages.

Code embeddings are also actively used in educational settings where
students use the same \llm-based tools to solve course
exercises~\cite{Denny2023Conversing,Yabaku2024University}. The tools
can also assist teachers in generating
exercises~\cite{Speth2023Investigating}, grading student
submissions~\cite{Balse2023Investigating}, or be used to track
students’ exercise progress~\cite{Paassen2021Mapping}.
Again, these approaches mainly focus on applications to textual
programming languages.

\begin{figure}[t]
  \begin{subcaptionblock}{0.4\textwidth}
    \centering \includegraphics[width=0.5\linewidth]{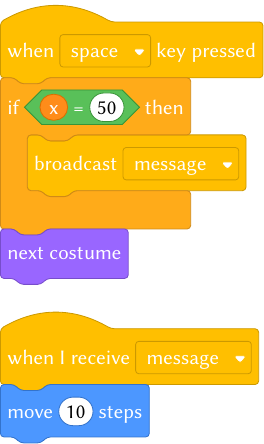}
    \caption{\label{fig:scratch-example}
      \Scratch code containing two scripts that are triggered by different events.
    }
  \end{subcaptionblock}
  \hfill \begin{subcaptionblock}{0.55\textwidth}
    \centering \includegraphics[width=\linewidth]{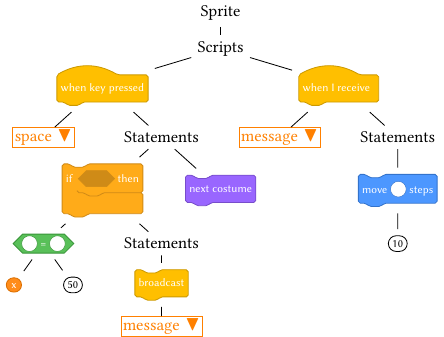}
    \caption{\label{fig:scratch-ast-example}
      The \AST contains nodes with a direct counterpart in the assembled program and abstract nodes inserted by \LitterBox.
    }
  \end{subcaptionblock}
  \caption{Example of the \Scratch code as it appears inside a sprite and its abstract syntax tree~(\AST).
    Each full \Scratch program can contain multiple such sprites that each contain a separate independent set of scripts.
  }
\end{figure}

However, especially in early software engineering education,
block-based visual programming environments such as
\Scratch~\cite{Maloney2010Scratch} or \snap~\cite{Harvey2010Bringing}
are frequently used instead~\cite{McGill2020Tools}. These environments
allow for the creation of programs by arranging pre-defined blocks
representing statements and expressions~(cf.\
\cref{fig:scratch-example}) on a two-dimensional working area. In
these environments, the assumption that code can be represented
similar to natural language text~\cite{Hindle2016naturalness} might
not hold since the language consists of freely placeable visual blocks
rather than a linear sequence of textual tokens.
Similarly, due to this different code representation, existing code
embedding models cannot be employed without prior adaptation. Some
models like for example \codeToVec~\cite{Alon2019Code2vec},
\codeToSeq~\cite{Alon2018Code2seq}, or \astnn~\cite{Zhang2019Novel} do
not use the text sequence directly, however, but use inputs based on
the abstract syntax tree~(\AST) of the code. \Scratch code can also be
parsed into such a format~(cf.~\cref{fig:scratch-ast-example}), which
suggests that the adaptation of existing models to the new domain is
at least technically feasible.

In addition to these foundational differences in language structure,
the type of programs found in \Scratch also differs. \Scratch is
mainly designed to implement small animations or interactive games
rather than data-manipulating algorithms. For this purpose, \Scratch
heavily relies on the concurrent execution of code in multiple
animated figures~(called sprites). \Cref{fig:scratch-example} shows a
typical example where the upper script executes whenever the user
presses the space key, and triggers the execution of further code by
broadcasting a message event.
Since programs written in regular programming languages rarely use
such event-handling-based code, common code embedding models rely on a
program’s syntactical structure to represent the
code~\cite{Alon2018Code2seq,Alon2019Code2vec,Zhang2019Novel,Ahmad2020Transformer}.
However, this might not be sufficient to capture such common \Scratch
language constructs, since the relations between event emitters and
receivers are lost. Thus, models additionally capturing control- and
data-flow information~(\eg~\ggnn~\cite{Allamanis2018Learning}) might
exhibit better performance on \Scratch-based tasks.

The \Scratch environment imposes additional natural-language
challenges not faced in most text-based programming scenarios.
The \Scratch user-base consists of large international communities for
example in Brazil~(\(\approx \perc{6.42}\) of all users),
Spain~(\(\approx \perc{3.33}\)), Poland~(\(\approx \perc{3.31}\)), or
China~(\(\approx
\perc{2.55}\)).\footnote{\url{https://scratch.mit.edu/statistics/},
accessed 2026-05-19} The English-default assumption for keywords and
method names in the libraries of many programming languages can be
difficult for international learners since they will have to
understand new-to-them programming concepts in an unfamiliar natural
language~\cite{Swidan2023Framework}. The \Scratch environment actively
tries to mitigate this by providing translations for the user
interface and all blocks. Additionally, custom variables, methods, and
sprite names are allowed to contain arbitrary Unicode characters
rather than being limited to alphanumeric characters.
This is stark contrast to the publicly shared open-source code found
for example on \github. In the \properNoun{Stack} code dataset mined
from open sources and created for the training of code-\llm{}s,
\perc{94} of the analysed \python files use English identifiers,
string constants, and documentation~\cite{Kocetkov2022Stack}.
Existing code embedding model approaches might therefore not only have
to be adapted to the different program structure of \Scratch, but also
need to take the differences in the natural languages used by
programmers into account.

While other work has already started using \llm{}s for
\Scratch-related
tasks~\cite{Druga2025Scratch,Fein2025LitterBox+,Fein2026Reasoning}, to
the best of our knowledge there so far exists no structured comparison
of the performance of different code embedding models for \Scratch.
This might be a result of many established code embedding comparison
tasks not being applicable to \Scratch. For example, the commonly used
benchmark of method-level code documentation generation or
summarisation~\cite{Alon2018Code2seq,Ahmad2020Transformer,Wang2021CodeT5,Sun2025Source}
is not feasible in \Scratch, since the projects do not contain any
documentation and only rarely contain code comments. Thus, creating suitable
training datasets is difficult. Similar challenges arise for the
commonly used method or variable naming
tasks~\cite{Allamanis2018Learning,Alon2018Code2seq,Alon2019Code2vec,Ma2022GraphCode2Vec},
since both programming concepts are only rarely used in publicly
available \Scratch projects~\cite{Aivaloglou2016How}.
However, the structure of \Scratch projects allows for the
construction of an alternative task. \Scratch projects consist of
multiple figures~(sprites) which can be animated using the code
contained within. Similar to classes in object-oriented programming
languages, each sprite therefore encapsulates its own state and
behaviour. Since all sprites in \Scratch must be named by the
programmer, large model training datasets can be created by mining
publicly shared \Scratch projects. To facilitate a baseline comparison
of code embedding models for \Scratch, we thus introduce the
\enquote{sprite naming} task in which the models must suggest suitable
sprite names given the code contained within.

Based on this foundational evaluation investigating whether
established code embedding models can be applied to the \Scratch
context, we evaluate potential applications of the models for
education-focussed applications, such as the prediction of program
correctness or estimation of exercise progression.
This paper extends our prior work~\cite{Fein2022Evaluation} which
evaluated the applicability of only the \codeToVec embedding
model~\cite{Alon2019Code2vec} to \Scratch.
In detail, the contributions of this paper are as follows:

\begin{itemize}[leftmargin=*]
\item We train five different code embedding models using various
  amounts of structural or control-flow information about the program
  on the sprite naming task to evaluate their applicability to
  \Scratch. We evaluate their performance in comparison to four
  different pre-trained \llm{}s.
\item To evaluate how the natural language used by the programmer for
  identifier names and strings impacts the model performance in, we
  collect 10 \Scratch program datasets each using a different primary
  natural language. We compare the models’ performance on the sprite
  naming task.
\item We evaluate fine-tuning strategies that adapt pre-trained sprite
  naming models to be used as classifiers for whole programs. This
  evaluates whether the models can be adapted to scenarios where
  insufficient training data for dedicated training is available.
\item To investigate the expressiveness of the embeddings for
  educational tasks, we evaluate whether pre-trained sprite-naming
  embedding models and embedding-\llm{}s can be used to estimate
  program correctness and student exercise progression.
\item We make our preprocessing tool,
  \EmbeddedKittens\footnote{\url{https://github.com/se2p/LitterBox},
  version 1.12, licenced GPL-3.0-or-later}, publicly available under
  an open-source licence to support further research in this area, and
  to support the development of tools making use of \Scratch code
  embeddings.
\end{itemize}

Our experiments show that models including more structural information
about the program outperform ones using a simpler program
representation on the sprite naming task. As expected, additionally
including control- and data-flow information results in further
improvements. In our experiments, all specifically trained \Scratch
embedding models outperform the pre-trained \llm{}s.
All models perform considerably worse when evaluated on programs that
use non Latin-based character sets which demonstrates that the
internationality of the \Scratch community needs to be taken into
account when developing models for the \Scratch programming language.
The pre-trained sprite-embedding models can be used estimate the
program correctness and student exercise progression without requiring
further fine-tuning, thus highlighting their applicability to
classroom scenarios with limited availability of training data.

\section{Background}

\subsection{Scratch}

\Scratch~\cite{Maloney2010Scratch} having over 135 million registered
users\footnote{\url{https://scratch.mit.edu/statistics/}, accessed
2026-05-19} highlights its popularity as a programming language
amongst teachers and learners. Its block-based concept allows students
to construct programs by arranging blocks that represent statements
and expressions. The block shapes clearly indicate in which positions
they can be placed, and the \Scratch interface subsequently prevents
invalid block combinations.
\Scratch contains many standard programming constructs such as loops,
conditions~(cf.\ \cref{fig:scratch-example}), or variables. However,
most blocks focus on the manipulation of attributes of sprites, the
interaction of different sprites, or reactions to user inputs. The
sprites can thus be animated to program small interactive games.

Since \Scratch is fully web-based, it can be used without installation
through its official website. This website actively encourages the
sharing of programs to allow other users to play the games and inspect
their code to gather ideas for their own projects.
The website also provides a \rest-\api through which the publicly
shared programs can be programmatically accessed. Large datasets of
programs as required for the training of machine learning models can
be created by sampling from the over 164 million publicly shared
programs.

\subsection{Code Embeddings}

Code embedding models map source code into a dense vector space
representation that aims to capture both semantic and syntactic
aspects of the code. This information can then be decoded by the
models to assist in code-related tasks.
Earlier code embedding models were trained to solve specific tasks
like for example suggesting method~\cite{Alon2019Code2vec} or variable
names~\cite{Allamanis2018Learning} or generating suitable
documentation~\cite{Ahmad2020Transformer}.
Since source code follows similar patterns as natural
language~\cite{Hindle2016naturalness}, large language models~(\llm{}s)
can be trained on large datasets of mixed natural language and code.
More recently, the focus has thus shifted mostly away from
task-specific models and instead aims to use the general-purpose
nature of large models~(\eg \properNoun{GPT-5}~\cite{Singh2026OpenAI},
\properNoun{Claude
Opus~4.7}\footnote{\url{https://www.anthropic.com/system-cards},
accessed 2026-05-19}) to solve code-related tasks~\cite{Hou2024Large}.
For example, these models allow for the generation of new code from
natural language descriptions or
specifications~\cite{Ma2026How,Tian2026Aligning}, the summarisation or
explanation of code in natural language~\cite{Haldar2024Analyzing}, or
code-code-transformations like the generation of test suites for
existing code~\cite{Schaefer2024Empirical}.
To further improve the \llm{}s’ performance for code-related tasks,
various fine-tuned models like
\properNoun{StarCoder~2}~\cite{Lozhkov2024StarCoder},
\deepseekcoder~\cite{DeepSeekAI2024DeepSeek}, or
\properNoun{Devstral~2}\footnote{\url{https://mistral.ai/news/devstral-2-vibe-cli},
accessed 2026-05-19} have been developed.

Both pre-\llm code embedding models and \llm-based approaches have
also been used in educational settings.
Since semantically similar programs should be mapped to similar
vectors in embedding space, these vectors can be used directly to find
clusters of students with similar solution approaches and help
educators find outliers that did not yet solve the
task~\cite{Bazzocchi2020Analyzing,Paassen2021Mapping}. Using the code
embeddings to determine similarity, this approach can be extended to
apply human feedback given for one student’s solution also
automatically to a similar program of another
student~\cite{Piech2015Learning}.
The main focus of \llm-based code embeddings in education lies on
undergraduate-level Python and Java tasks~\cite{Raihan2025Large}.
\llm-based tools are supporting teachers when generating
exercises~\cite{Speth2023Investigating}, assessing student
submissions~\cite{Balse2023Investigating,GonzalezCalatayud2021Artificial},
or to generate next-step hints~\cite{Roest2024Next}. Students are
actively using \llm{}s to solve
exercises~\cite{Denny2023Conversing,Yabaku2024University}.

While \llm{}s have been used in the \Scratch context to generate
exercise material~\cite{Grassl2025Detecting}, the main focus seems to
be on supporting students in open-ended tasks with chat-based
interfaces integrated into the \Scratch environment. These can be used
to find inspirations for additional features of the
programs~\cite{Druga2023Scratch,Druga2025Scratch}, or to receive help
in locating and fixing issues in the program by receiving an
\llm-generated natural language description of the underlying
misconceptions and the issue’s impact for the specific
program~\cite{Fein2025LitterBox+,Fein2026Challenges}.
Apart from our previous study applying the \codeToVec model to
\Scratch~\cite{Fein2022Evaluation}, applying dedicated non-\llm-based
code embeddings to \Scratch programs to the best of our knowledge
remains an unexplored area.

\section{Transforming Scratch Code Into the Model Input Formats}

Code embedding machine learning models for regular programming
languages assume that the code exists in a textual format to which
further processing steps can be applied to obtain the model-specific
input format. Thus, before being able to apply the models to a visual
programming language, additional steps converting the visual
representation into a textual one are necessary first.
Specifically, we describe our preprocessing for the models used in our
experiments: Transformer models~\cite{Vaswani2017Attention} and
transformer-based \llm{}s use flat sequences of tokens to represent
the program, \codeToVec~\cite{Alon2019Code2vec} and
\codeToSeq~\cite{Alon2018Code2seq} use an abstraction derived from the
abstract syntax tree~(\AST) of the program as input,
\astnn~\cite{Zhang2019Novel} uses a sequence of subtrees of the \AST,
and \ggnn~\cite{Allamanis2018Learning} additionally requires an
extension of the \AST into a graph containing control and data flow
information.

\subsection{The Scratch Data
Format}\label{sec:code-processing:scratch}

The code of a \Scratch project is saved in the form of a single \json
file.
This file contains a list of all sprites of the project. For each
sprite its name, \enquote{costumes} (\ie images representing what the
figure looks like), sounds, and the actual code it contains, are
stored.
Within sprites, the code is structured into several
scripts~(cf.~\cref{fig:scratch-example} shows two scripts). However,
all code of a sprite is saved as a single flat list of all its blocks.
Each block is represented by a unique identifier and an
\enquote{opcode} which is a unique name that identifies which type of
block it is.
Since this storage format loses all information about the actual
syntactical nesting of the program, each block also contains
references to relevant related blocks using the unique block
identifier: its parent block, its input blocks (\eg~the left and right
operands of an addition block), and for statements also a pointer to
the next block attached to the bottom of it. For example, the
\inlinescratch{\blockevent{if \boolempty~then}} block in
\cref{fig:scratch-example} contains references to the
\booloperator{\ovalnum{~} = \ovalnum{~}}, the
\inlinescratch{\blockevent{broadcast}}, and the
\inlinescratch{\blocklook{next costume}} blocks.
Those pointers allow for the recreation of the nested block structure
when parsing the project’s \json file. The nesting of the visual
blocks behaves similar to the use of parentheses and braces in a
textual programming language.

We make use of \LitterBox~\cite{Fraser2021LitterBox} to parse \Scratch
programs into their abstract syntax tree (\AST)
representation~(cf.~\cref{fig:scratch-ast-example}). \LitterBox
supports parsing all basic \Scratch programs and can also parse
programs using additional blocks provided by some of the officially
supported extensions available in the \Scratch user
interface~(\emph{pen} to draw on the game canvas, \emph{music},
\emph{text to speech}, and \emph{translation}).
Since \Scratch is a block-based language, it does not allow for the
introduction of syntactical errors. Therefore, \LitterBox can
construct a valid \AST for all projects that only contain supported
blocks and have not been manually tampered with outside the \Scratch
user interface.

All the following model-specific processing steps are then implemented
on top of this \LitterBox-generated \AST as part of our
\EmbeddedKittens extension to \LitterBox. Both tools are available
publicly under an open-source licence at
\url{https://github.com/se2p/LitterBox}.

\begin{figure}[t]
  \centering 

  \begin{subcaptionblock}{0.6\textwidth}
    \begin{lstlisting}[aboveskip=0pt,belowskip=0pt]
BEGIN_SCRIPT
event_whenkeypressed key ( 32 )
control_if < ( x ) operator_equals ( 50 ) >
BEGIN_SUBSTACK
event_broadcast event_message ( message )
END_SUBSTACK
looks_nextcostume
END_SCRIPT

BEGIN_SCRIPT
event_whenbroadcastreceived event_message ( message )
motion_movesteps ( 10 )
END_SCRIPT
    \end{lstlisting}
    \caption{\label{fig:transformer-representation}\neuralcodesum transformer: Flat sequence of abstract tokens.}
  \end{subcaptionblock}
  \begin{subcaptionblock}{0.39\textwidth}
      \begin{lstlisting}[aboveskip=0pt,belowskip=0pt]
    when [space v] key pressed
    if <(x) = (50)> then
        broadcast [message v]
    end
    next costume

    when I receive [message v]
    move (10) steps
      \end{lstlisting}
    \caption{\label{fig:llm-representation}\scratchblocks format used for \llm{}s.}
  \end{subcaptionblock}

  \vspace{\baselineskip}

  \begin{subcaptionblock}{0.7\textwidth}
    \centering\small
    x (eq if stmts broadcast) message\\
    space (whenkeypressed stmts) next-costume\\
    \(\dots\)\\
    \(\phantom{\text{\rmfamily hash}}~\Downarrow~\text{\rmfamily hash}\)\\
    x (\texttt{-1654711692}) message\\
    space (\texttt{307684711}) next-costume\\
    \(\dots\)
    \caption{\label{fig:code-to-vec-representation}
      \codeToSeq~(top) and \codeToVec~(bottom):
      Paths between leaves of the \AST.\@
      For \codeToVec the sequence of non-leaf tokens on the path is hashed.
    }
  \end{subcaptionblock}

  \vspace{\baselineskip}

  \begin{subcaptionblock}{0.7\textwidth}
    \includegraphics[width=\linewidth]{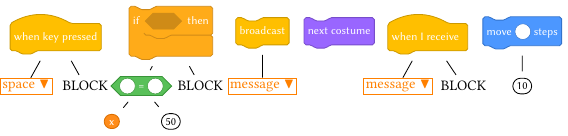}
    \caption{\label{fig:astnn-representation}\astnn: \AST split into a sequence of statement-\AST{}s.
      The \emph{BLOCK} marker nodes indicate that a nested statement list was cut off from the statement-tree at this location.
    }
  \end{subcaptionblock}
  \caption{\label{fig:scratch-representations}Processing steps for the program from \cref{fig:scratch-ast-example} to be usable as input to \codeToSeq, \codeToVec, \astnn, \neuralcodesum, and \LLMs.
  }
\end{figure}
 
\subsection{Dedicated Token-Based
Models}\label{sec:code-processing:tokens}

For this category of models, represented by the
transformer-based \neuralcodesum~\cite{Ahmad2020Transformer} in our
experiments, the code is represented in a flat textual format. For
regular programming languages, this simplifies the application of such
models since the source code already is in the required format.
For the visual language \Scratch, neither the format as shown to the
user, nor the \AST can be directly used. Instead, we designed a
translation step from the \AST into a textual representation.
The translation process is implemented by walking over the \AST in
depth-first traversal and appending a textual representation for each
encountered node to the output text stream. While the \Scratch user
interface imposes no fixed ordering between scripts placed on the
two-dimensional workspace, the \AST traversal implicitly serialises
the scripts in the into the order in which their first blocks are
appearing in the project’s \json. For each sprite, the list of blocks
in the \json represents the order in which the blocks were added to
the project.
We retain some structural information in the text similar to the
braces in some programming languages~(\eg~Java, C). We mark the
beginning of scripts and indented block stacks inside control
structures with special \verb|BEGIN_SCRIPT| and \verb|BEGIN_SUBSTACK|
tokens respectively and also insert corresponding \verb|END_…|
markers~(cf.\ \cref{fig:transformer-representation}).

\Scratch visually categorises statements of different categories by
colouring them. For example, all statements relating to the movement
of a sprite are coloured blue.
In the textual representation we retain this information by not using
the English block text as shown to the user, but by instead converting
it to an abstract identifier using the group as prefix and a single
word for the concrete block name. Expressions used as inputs to blocks
are enclosed with parentheses to mimic function calls of regular
programming languages. For example, the textual representation of a
\inlinescratch{\blockmove{move \ovalnum{10} steps}} statement is
\verb|motion_movesteps (10)|. For boolean inputs, angles~(\verb|<|,
\verb|>|) are used instead. Both cases mimic the shape used in the
visual representation and potentially aid the model in learning from
the surrounding context.

In case blocks have multiple inputs, we tokenise them in the
left-to-right reading order as they appear in the English variant of
the block. For example, \ovaloperator{letter \ovalnum{1} of
\ovalnum{apple}} results in the tokenisation
\verb|operator_letterof (1) (apple)|.
However, \Scratch adapts the order of the inputs to embed them as far
as possible into the natural grammatical structure of the block text
in the user’s language settings. When changing the language to
Hungarian for the previous example, the block is shown as
\ovaloperator{\ovalnum{apple}~\ovalnum{1} bet\H{u}je}. In cases where
the model will be trained for example on a dataset obtained in a
classroom setting where it is known that most students use the local
language in the \Scratch user interface, the tokensation order might
have to be adapted accordingly.

Contrary to this tokenisation of statements in Polish
notation~(\ie~using the operator as prefix with following operands),
we tokenise mathematical and logical operators like \(+\), \(=\), or
\texttt{and} in their natural-language-independent infix order as also
used by most regular textual programming languages.
Finally, variable names, string literals, and numbers are kept as-is
without changes. This allows the model-specific text-encoder to treat
them as required without losing information beforehand during the
transformation from \AST to text.

\subsection{Large Language Models}\label{sec:code-processing:llms}

Since \LLMs are Transformer-based, they also receive flat token
sequences as inputs. However, unlike dedicated models, they have been
pre-trained on a large corpus of textual resources already. The unseen
specially crafted token representation described in the previous
section might therefore not be suitable for \LLMs. Instead, the
\scratchblocks
representation\footnote{\url{https://en.scratch-wiki.info/wiki/Block_Plugin/Syntax},
accessed 2026-06-05} as shown in \cref{fig:llm-representation} has
likely been part of the training dataset of \LLMs due to its use for
example on the \Scratch community
forums.\footnote{\url{https://scratch.mit.edu/discuss/}, accessed
2026-06-05}
Prior research confirms that this format works better than the native
\json program representation when prompting
\LLMs~\cite{Fein2026Reasoning}, and the \scratchblocks representation
has also been adopted successfully in other research prompting \LLMs
with \Scratch code~\cite{Fein2025LitterBox+,Grassl2025Detecting}.
Thus, we use the \Scratch code to \scratchblocks conversion as already
built into \LitterBox and adopt this text representation when
prompting \LLMs in our experiments.
Since the conversion to \scratchblocks is internally implemented as an
\AST traversal, the scripts are again serialised in the order they
were added to the project~(cf.~\cref{sec:code-processing:tokens}).

\subsection{Code2vec and Code2seq}\label{sec:code-processing:code2vec}

The \codeToVec~\cite{Alon2019Code2vec} and
\codeToSeq~\cite{Alon2018Code2seq} embedding models are based around
the idea of connecting leaf nodes of the abstract syntax tree and
representing a code snippet as its set of all possible such
connections~\cite{Alon2019Code2vec,Alon2018Code2seq}. This means that
a piece of code is represented by its set of possible connections
between leaves of its \AST. To limit the computational effort, a
length threshold model hyperparameter can be used to prune long
connections.
During model training an embedding vector for each such connection is
learned, and the final code is represented as a weighted average over
these embeddings~\cite{Alon2019Code2vec}.
The \codeToSeq model retains the sequence of individual nodes that are
visited when walking the tree between the two chosen
leaves~(cf.~\cref{fig:code-to-vec-representation}). In case of
\codeToVec, the path between the leaves is saved as only a hash value
of its string representation.
After obtaining the \AST from \LitterBox, the set of connections can
be constructed like on the \AST of a regular programming language by
traversing the \AST.

A difficulty for this representation of the code as paths comes from
the structure of sprites. Since they consist of multiple independent
scripts, each of those scripts results in a subtree of the \AST that
is only connected to the other scripts at the top via a virtual
\enquote{ScriptList} node which acts as a parent for all scripts of
the sprite.
Therefore, choosing a short path length during preprocessing results
in few connections between scripts but instead results in most paths
representing connections within scripts. A longer path length would
allow the model to capture more information across scripts later-on.
However, such an increase of the path length also has downsides:
Firstly, it considerably increases the time required to preprocess the
code. Additionally, since it is not possible to give the full set of
paths to the model during training due to \gpu memory constraints, a
fixed number of random samples~(200 in case of the original \codeToVec
experiments on Java code~\cite{Alon2019Code2vec}) has to be selected
for each sprite. This therefore decreases the chance of selecting the
ones connecting leaves significant to the sprite’s functionality~(\eg
the \enquote{when this sprite clicked} event for a button).

\subsection{ASTNN\@: AST-based Neural Network}

The \astnn model~\cite{Zhang2019Novel} incrementally combines
information from child nodes into their parents starting from the
leaves of the \AST.\@ To make this process more efficient and reduce
the amount of lost information for deeply nested
\AST{}s~\cite{Alon2020Bottleneck}, the \AST is split into a sequence
of separate per-statement subtrees. The individual subtree embeddings
are combined into a single embedding using a bidirectional gated
recurrent unit~(\gru) encoder~\cite{Zhang2019Novel}.

The per-statement subtrees of the original \AST~(cf.\
\cref{fig:scratch-ast-example}) are shown in
\cref{fig:astnn-representation}.
In \Scratch, each script starts with an initial event statement at the
top~(\eg the \enquote{when I receive message} block as shown in
\cref{fig:scratch-example}). Its bottom has an indented marker onto
which a further regular statement can be attached.
A special case are statements representing loops and conditional
control structures. These do not only contain a marker at the bottom,
but also allow for the connection of additional statements
inside~(\eg\ the \inlinescratch{\blockevent{if \boolempty{} then}} in
\cref{fig:scratch-example}). The control structure in this case
extends around the blocks inside the scope of the loop or condition.
In the subtree given to \astnn, such inner blocks are represented by
an abstract \verb|BLOCK| node.
Inputs to statements, \ie expressions, have a distinctly different
shape, being either rounded for numeric and textual inputs, or angled
for boolean conditions.
This clear visual separations of statements, their grouping into
enclosed blocks, and expressions is also retained in the \AST as
easily separable subtrees per statement with all input expressions
being part of the same subtree. All statements that are part of the
same indentation level are always children of the same virtual
\enquote{StmtList} node, thus also clearly marked.
Therefore, the splitting of the \AST into subtrees per statement can
be performed similar to how it can be done for a textual programming
language that uses semicolons as statement separator and braces for
the scoping of statements into control structures.

\subsection{GGNN\@: Gated Graph Neural Network}

\begin{figure}[t]
  \centering \begin{minipage}{0.2\textwidth}
    \begin{scratch}
      \blockvariable{set \selectmenu{var} to \ovalnum{0}}
      \blockif{if \booloperator{\ovalvariable{var} > \ovalnum{-1}} then}{\blockevent{broadcast \selectmenu{message}}
      \blockvariable{change \selectmenu{var} by \ovalnum{1}}
      }
      \blockspace \blockinit{when I receive \selectmenu{message}}
      \blocklook{say \ovalnum{Hello}}
    \end{scratch}
  \end{minipage}
  \hspace*{1em}
  \begin{minipage}{0.68\textwidth}
    \includegraphics[width=\linewidth]{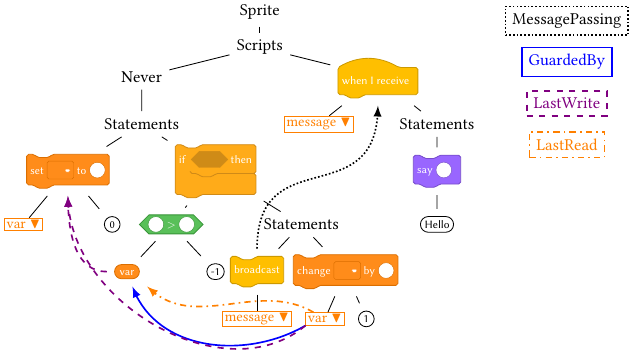}
  \end{minipage}
  \caption{\label{fig:ggnn-representation}Example code and its representation for \ggnn as
    graph based on the \AST with additional edges for control and data flow information.
    LastLexicalUse edges are not shown since the relevant connections are already made by LastWrite and LastRead edges.
    Edges are shown directionally to highlight their construction.
  }
\end{figure}
 
As a graph neural network, the \ggnn
model~\cite{Allamanis2018Learning} receives a graph as input. This
graph structure is retained internally while training the model. The
graph is based on the \AST and labels each node with a semantic
embedding. During training, multiple rounds of information exchange
between nodes along the graph edges propagate this information and
adapt the embedding vectors of neighbouring nodes accordingly. A
special node is added to the graph such that its final embedding is
used as representation for the whole
input~\cite{Allamanis2018Learning}.
Thus, the preprocessing step aims to construct a program graph with
semantically relevant edges between nodes.

We construct the context graph that contains the syntactic and
semantic relationships between the \AST nodes following the original
approach used on Java code~\cite{Allamanis2018Learning}.
The node set of the graph and its initial edge set are given by the
\AST. Each node is labelled by the type of the node~(\eg IfThen), or
in case of nodes that contain user-defined values~(\eg literal values,
or the name of a variable) a normalised string representation of this
value.
We also connect each node to its syntactical successor. In case of a
parameterless statement, this is the statement attached to the bottom
of it. Similar to the tokenisation described in
\cref{sec:code-processing:tokens}, we again follow the left-to-right
reading order in the English variants of the blocks for blocks that
accept multiple inputs.
For example, \ovaloperator{letter \ovalnum{1} of \ovalnum{apple}}
results in the connections \ovaloperator{letter
of}~\(\rightarrow\)~\ovalnum{1}~\(\rightarrow\)~\ovalnum{apple}.

To construct further semantic edges, we again use available \LitterBox
features to extract the control and data flow dependencies within the
program. For the data flow analysis, we consider both user-defined
variables and sprite attributes that can be changed by the user (\eg
size, x/y-position) since both can be accessed via expression blocks.
When referring to variables in the following definitions we consider
both variables and attributes. Following the edge nomenclature of
\citeauthor{Allamanis2018Learning}~\cite{Allamanis2018Learning} we
construct the semantic connections of the program
graph~(cf.~\cref{fig:ggnn-representation}) as follows:

\begin{description}

  \item[LastRead, LastWrite] Such edges are added between each
    variable use and the previous time it was accessed. These
    connections can be directly extracted from the data dependency
    graph.

  \item[LastLexicalUse] This type of edge connects variables to the
    last position in the program \enquote{text} where they appeared
    previously independent of control flow. We only consider last uses
    within scripts or procedures since there is no strict order
    between different scripts. In case the same variable is used
    multiple times inside a single block, we again use the block order
    as it appears in the English variant.

  \item[ComputedFrom] These edges connect all variables appearing on
    the right side of an assignment to the variable on the left side.

  \item[GuardedBy, GuardedByNegation] This type of edge connects each
    variable appearing in the condition of a control statement (if,
    if-else, loops) to all uses of the same variable in blocks which
    are enclosed by the control statement.

  \item[FormalArgName] Custom procedures in \Scratch can have
    parameters. We connect their declaration to all their uses within
    the custom procedure.

  \item[ReturnsTo] Custom procedures in \Scratch cannot return values.
    Therefore, we only connect the last statement in a procedure back
    to its declaration.

\end{description}

Additionally, we construct semantic edges specific to \Scratch
programs.
Each sprite in \Scratch can contain multiple scripts that are executed
concurrently. A script can either start executing directly on program
start, or alternatively be triggered later by user inputs or other
state changes in the program.
\Scratch also allows to programmatically trigger the execution of
other scripts by broadcasting messages~(cf.\ \enquote{broadcast} block
in \cref{fig:scratch-example}). For these cases we add additional
edges to the \ggnn context graph for all corresponding sending and
receiving block pairs.
Similarly, for custom procedures we connect the calling block to the
definition. In case the procedure has parameters, we connect each of
the values passed in the caller to their corresponding definition in
the procedure instead.
Finally, we add all backwards edges for all edge types to the
constructed graph. This speeds up the propagation of information
through the network~\cite{Allamanis2018Learning}.

\section{Experiments}

To provide a better understanding of code embeddings for \Scratch
programs, we first need to evaluate whether existing code embedding
models can be applied to the new domain. However, common model
evaluation tasks such as documentation
generation~\cite{Alon2018Code2seq,Ahmad2020Transformer,Wang2021CodeT5,Sun2025Source}
or method
naming~\cite{Alon2018Code2seq,Alon2019Code2vec,Ma2022GraphCode2Vec} are not
possible in \Scratch due to lack of suitable training
data~\cite{Aivaloglou2016How}. Instead, we introduce the
\enquote{sprite naming} task as an alternative code summarisation
task. Similar to method naming, the models are tasked to suggest a
suitable sprite name based on the code contained within. To establish
a baseline, we compare multiple code embedding models using this task
as part of our first research question:

\begin{description}
\item[RQ1] Can code embeddings be used to summarise sprites, \ie be
    used to predict their name?
\end{description}

The \Scratch user base predominantly consists of an international
community of young learners. Since these learners might not yet be
familiar with English, the \Scratch user interface can be switched to
alternative languages and allows arbitrary Unicode characters for
sprite names. Consequently, the programmers likely use their familiar
natural language to name variables or sprites, which directly elicits
our second research question:

\begin{description}
\item[RQ2] How does the internationality of \Scratch influence the
    sprite name prediction?
\end{description}

The internal code embeddings of the models trained on the sprite
naming task might be overly specific for this application. To
investigate whether the embeddings are sufficiently expressive and
contain generalising semantic information, we evaluate whether the
pre-trained models can be fine-tuned for program classification tasks:

\begin{description}
\item[RQ3] What is the best method to fine-tune the sprite
    embeddings for program classification?
\end{description}

The sprite naming task mainly serves as a baseline comparison for the
model performance, but it is unlikely to be directly relevant for
educational scenarios. In our final two research questions, we
therefore investigate whether the code embeddings are sufficiently
expressive to support more education-focussed tasks, such as
predicting program correctness or exercise progress:

\begin{description}
\item[RQ4] Can \Scratch program embeddings be used as surrogate model
    for program correctness?
\item[RQ5] Can \Scratch program embeddings be used to determine
    student exercise progress?
\end{description}

\subsection{Dataset Collection}\label{sec:setup:dataset-collection}

To create sufficiently sized datasets for model training, we used
datasets consisting of projects which have been publicly shared on the
\Scratch website. As per the terms of
use,\footnote{\url{https://scratch.mit.edu/terms_of_use} Section 4.3,
accessed 2026-01-15} all such projects are shared under a Creative
Commons CC-BY-SA-2.0 licence explicitly allowing the re-use of the
programs under the same conditions as long as attribution to the
original author is
given.\footnote{\url{https://creativecommons.org/licenses/by-sa/2.0/},
accessed 2026-01-15}
The \Scratch website actively encourages such re-use and adaptation by
providing a \enquote{remix} functionality similar to the
\enquote{fork} mechanism employed by the open-source community on
\properNoun{GitHub}. The feature allows users to start a new project
by using any existing project as starting template instead of starting
from an empty one.
The \Scratch \api provides the metadata required to extract the
ancestor relationship between base and remixed projects. We ensured
that our datasets do not contain both parent and remix projects to
avoid a potential data leakage between training and evaluation splits
when creating datasets for the purpose of training our machine
learning models.

\subsection{Model Implementations}\label{sec:model-implementations}

We trained most models specifically for our experiments. For \llm{}s,
we used their pre-trained versions available via the respective
\api{}s without further fine-tuning instead.

\subsubsection{Task-Specific Training: NeuralCodeSum, code2vec,
code2seq, ASTNN, GGNN}

For all models we trained ourselves, we used the same model
implementations that can be found in the replication package of a
prior publication~\cite{Fein2026Challenges}. The replication package
contains implementations for the \codeToVec~\cite{Alon2019Code2vec},
\codeToSeq~\cite{Alon2018Code2seq}, \astnn~\cite{Zhang2019Novel},
\ggnn~\cite{Allamanis2018Learning} models, and for the
\neuralcodesum~\cite{Ahmad2020Transformer} transformer.
These models were used to predict method names in a Java dataset,
which closely matches the sprite naming task of our RQ1.
The \neuralcodesum transformer was originally designed for the
summarisation of code into a brief natural language
sentence~\cite{Ahmad2020Transformer}. Unlike other general-purpose
transformer models, it has therefore been designed specifically for
code-related tasks. While newer or larger transformer-based models are
available, most \Scratch projects in our training datasets are fairly
small and do not require large input context window sizes. The
generation of long output sequences is also not required for our
sprite naming task.

We designed our data preprocessing tool to output data in the same
format as already used by the model implementations. Thus, no further
code changes to the models themselves were required.
The \astnn and \codeToVec models represent the output as a single
label rather than a sequence of subtokens. For the other models, a
beam search decoder is used when predicting textual sequences.

\subsubsection{Pre-Trained: Large Language Models}

We used \properNoun{GPT-5-mini}~(version: gpt-5-mini-2025-08-07) as
available via the \properNoun{OpenAI} \api. While the full
\properNoun{GPT-5} model would provide more advanced reasoning
features~\cite{Singh2026OpenAI}, we did not require these for our
experiments. All prompts have been submitted to the \api in February
2026.
Additionally, we set up a self-hosted instance of \ollama{} to host
three additional text-generating \LLMs:
\deepseekcoder~\cite{DeepSeekAI2024DeepSeek},
\mistral,\footnote{\url{https://mistral.ai/news/mistral-small-3-1},
accessed 2026-06-05} and \gemma~\cite{GemmaTeam2025Gemma}.
For the later research questions~(RQs 3--5) that require embeddings
rather than generated text, we also self-hosted the
\qwenembedding~\cite{Zhang2025Qwen3} embedding
\llm.\footnote{\deepseekcoder:
  \url{https://ollama.com/library/deepseek-coder-v2:16b}, hash
  63fb193b3a9b; \mistral:
  \url{https://ollama.com/library/mistral-small3.1:24b}, hash
  b9aaf0c2586a; \gemma:
  \url{https://ollama.com/library/gemma3:27b-it-qat}, hash
29eb0b9aeda3; \qwenembedding
\url{https://ollama.com/library/qwen3-embedding:8b}, hash
64b933495768}
We chose \deepseekcoder, \gemma and \mistral to compare one dedicated
code model and two state-of-the-art general purpose models that are
self-hostable on a single \gpu. Prior research has shown that \gemma
is suitable for \Scratch-related tasks~\cite{Fein2026Reasoning}.
While for \gemma the dedicated embedding variant
\embeddinggemma~\cite{Vera2025EmbeddingGemma} exists, its short
context window of only \exnum{2048} tokens would truncate many
\scratchblocks input sequences. We therefore chose \qwenembedding as
alternative due to its performance in the \initialism{mteb}
benchmark~\cite{Enevoldsen2025MMTEB}.
By setting the maximum input context size of \ollama to \exnum{16384}
tokens, only 32 samples were truncated when prompting the \LLMs for
the sprite naming task.
Since we could self-host most of the \LLMs used for our experiments
and thus did not have to pay to access their \api{}s, only the usage
of the \properNoun{OpenAI} \api resulted in total costs of
approximately 100\,USD\@. Using a larger \properNoun{GPT}
model~(\eg~\properNoun{GPT-5.4}) would have increased costs by a
factor of seven.

Even though all these \LLMs are likely pre-trained on data crawled
from the official \Scratch community forums, we believe the risk of a
data leakage to be small since users in the forum likely refer to
smaller code snippets in \scratchblocks format as part of discussions
or questions about concrete parts of the code rather than putting
large code snippets into their forum posts.
Whole \Scratch projects are also unlikely to be part of the training
data, since downloading the original program-\json from the \Scratch
\api is a rate-limited two-step process, the result of which would
then have to be converted to \scratchblocks format~(\eg using
\LitterBox) to be similar to our inputs. This is unlikely to be
implemented in a general-purpose web crawler.

\subsection{RQ1: Using Code Embeddings for Code Summarisation}

\subsubsection{Dataset}\label{sec:setup:sprite-naming-dataset}

To train the models for the task of naming sprites, we randomly
sampled \numProjectsSpriteNaming publicly shared \Scratch projects.
After removing projects which \LitterBox could not process, we split
the projects in an 80:10:10 split into training, validation, and
testing subsets. Since we want the model to learn to generate
user-chosen names, all sprites having a default name~(\eg
\enquote{Sprite 1} for English- or \enquote{Hahmo1} for
Finnish-speaking users) were ignored. Similarly, while the background
image is modelled as \enquote{Stage} and can contain code, it cannot
be renamed by users and was therefore also excluded. We retained
sprites with names also present in the sprite catalogue, since the
user still actively selected a suitable image and thereby its
corresponding predefined name~(\eg~\enquote{Hedgehog},
\enquote{Rocketship}).
Finally, sprites not containing any code were removed.
This resulted in \numSpritesSpriteNamingTrain sprites in the training,
\numSpritesSpriteNamingEval in the validation, and
\numSpritesSpriteNamingTest in the test dataset.

\begin{figure}[t]
  \centering \includegraphics[width=\textwidth]{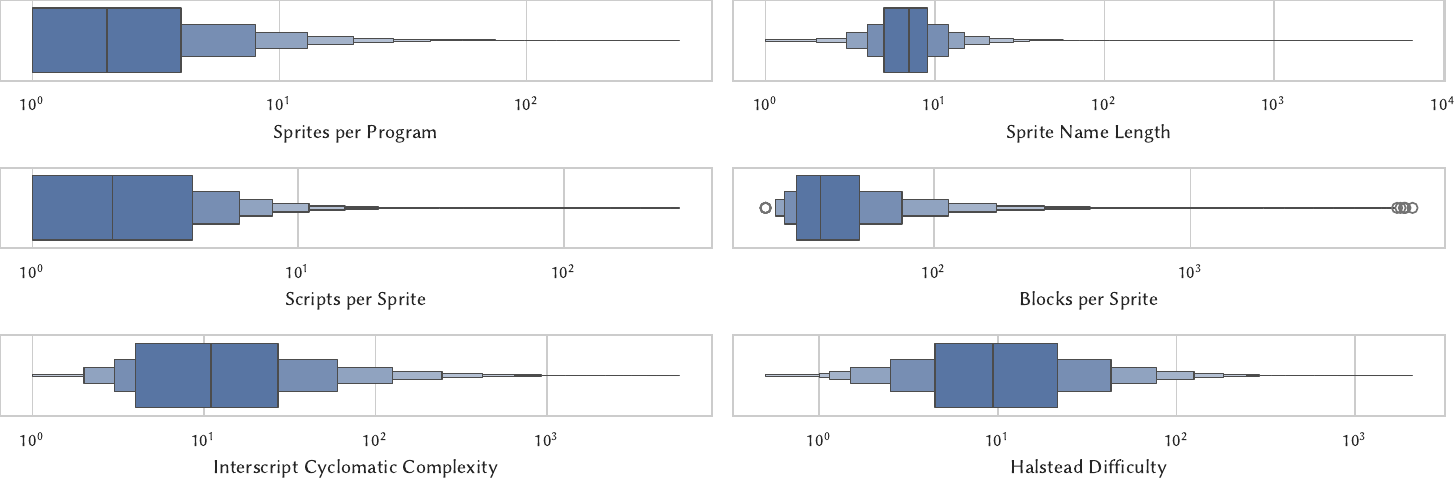}
  \caption{\label{fig:sprite-naming-dataset-stats}Sprite naming dataset structure.}
\end{figure}
 
As shown in \cref{fig:sprite-naming-dataset-stats}, most programs
contain between 1 and 4 sprites~(median \(\md = 2\), interquartile
range \(\iqr=3\)) sprites relevant for the sprite-naming task. These
sprites usually have short names~(\(\md=7\), \(\iqr=4\) characters)
and are each organised into few separate scripts~(\(\md=2\),
\(\iqr=3\)). Nevertheless, most sprites contain a meaningful number of
blocks~(\(\md=36\), \(\iqr=22\)).
As the complexity metrics computed by \LitterBox show, most sprites
contain actual logic making use of branching constructs and operators
rather than being a linear sequence of statements~(cyclomatic
complexity~\cite{McCabe1976Complexity}: \(\md=11\), \(\iqr=23\),
Halstead difficulty~\cite{Halstead1979Elements}: \(\md=\num{9.375}\),
\(\iqr=\num{17.192}\)).

Unlike identifiers in most programming languages, \Scratch sprite
names can contain arbitrary Unicode text. To reduce the vocabulary
sizes required in the machine learning models, we subtokenised and
normalised the sprite names. We subtokenised the names by splitting on
camel case boundaries, boundaries between letters and digits,
whitespace, and punctuation characters. The subtokens were then
normalised by removing punctuation and converting the characters to
lowercase. Characters of writing systems which do not differentiate
between upper- and lowercase variants remained unchanged. For
embedding models that do not model the labels as subtokens but as
single whole tokens~(\ie for \astnn and \codeToVec), we concatenated
the normalised subtokens back together with \enquote{|} characters in
between.

\subsubsection{Model Training and Hyperparameter Search}

\begin{figure}[t]
  \lstset{columns=fullflexible,breakindent=0pt}
  \begin{lstlisting}
The code below is written in the Scratch programming language. It is in the ScratchBlocks format which you know for example from the Scratch community forums. The code belongs to a single sprite in the program. I want to find a suitable name for this sprite. Suggest 10 names for the sprite. Only return the list of names with one name per line.
```
{{ source_code }}
```\end{lstlisting}
  \caption{\label{fig:sprite-naming-prompt}\llm prompt used for the sprite naming task. \texttt{source\_code} is replaced with the sprite’s code in \scratchblocks format.}
\end{figure}
 
We trained all non-pretrained models~(cf.\
\cref{sec:model-implementations}) on our sprite naming dataset. For
each model, we used the training checkpoint with the lowest validation
loss as the model for later evaluation.
Since the \LLMs are already pre-trained, no further training was
required. Instead, we prompted the models~(cf.\
\cref{fig:sprite-naming-prompt}) via the official \api or the
self-hosted \ollama endpoint and requested them to return ten
suggestions for suitable names for the sprite. The \Scratch code was
integrated into the prompt in the \scratchblocks format. In most cases
the models responded with an itemised list we could extract the first
ten names from. When the model generated fewer than ten suggestions,
we excluded the sample from our evaluation to later be able to compute
the top-10-accuracy. This filtering step removed between
\exnum{4}~(\properNoun{GPT-5-mini}) and \exnum{1269}~(\perc{0.8},
\mistral) responses.

We focussed our hyperparameter search mainly around parameters that
influence the structure of the input or the model itself. Since a full
grid-search would have been too computationally expensive, we
iteratively fixed all but one hyperparameter in each training run to
narrow down the search space.
For \astnn, \codeToVec, and \ggnn, an embedding size of \exnum{128}
performed best while larger embedding sizes of \exnum{256} and
\exnum{512} yielded the best results for \codeToSeq and
\neuralcodesum, respectively~(search space for all models:
\exnum{128}, \exnum{256}, and \exnum{512}).
For \codeToVec and \codeToSeq, increasing the \AST path
length~(cf.~\cref{sec:code-processing:code2vec}) to larger values than
8 did not improve the results. We included path lengths up to 12 in
our parameter search. Increasing the lengths further would have been
too computationally expensive in the data preprocessing stage. The
same length of 8 was also originally used on Java
code~\cite{Alon2018Code2seq,Alon2019Code2vec,Fein2026Challenges}.
Since using all possible \AST paths as model input would be too
computationally expensive, 200 random ones are sampled for each
sprite. Neither halving nor doubling the number of paths improved the
results, even when an increased maximum path length resulted in a
larger number of paths per sprite.
For \astnn, we excluded all programs with a tree depth greater than 30
from the training dataset to limit the required computational effort
during model training.
We configured \neuralcodesum to use 6 Transformer layers~(search
space: 4--12). Our \ggnn model uses 12 graph convolution layers as
encoder~(search space: 4, 8, 12, 16). \ggnn uses four \lstm layers as
decoder from embeddings into token sequences while \codeToSeq uses a
single \lstm decoder layer~(search space: 1, 2, 4, 6).

\subsubsection{Evaluation}\label{sec:rq1:methods-eval}

To evaluate the prediction performance, we used established machine
learning metrics. We computed the top-\(k\)-accuracy~(\ie one
perfectly matching name suggestion within the top-\(k\) predicted
names) and the F1-score~(macro-F1 for perfectly matching names) to
compare whole predicted to original names.
For a more fine-granular evaluation estimating how close the model
prediction are to the original name even when not being a perfect
match, we also considered multiple sub-token-based metrics. We report
their mean values over the whole dataset as the respective overall
score.
The \bleu score~\cite{Papineni2001BLEU} takes into account how many of
the subtokens of the original name also appear in the prediction. For
\rouge~\cite{Lin2004ROUGE}, we report the \rouge-2-F1-score to additionally
take into account how often bi\-grams of subtokens are correctly
predicted and thus estimate how frequently the correct subtokens not
only appear in the predicted name but also are in the correct order.
Finally, we compute the \meteor score~\cite{Banerjee2005METEOR} as
implemented by the \properNoun{NLTK}
library.\footnote{\url{https://www.nltk.org/}, accessed 2026-05-19}
Since this implementation assumes an English text, advanced \meteor
score features like stemming and ignoring synonyms as defined by the
\properNoun{WordNet}~\cite{Miller1995WordNet} corpus might not be
accurate for all data-points in our multilingual \Scratch dataset.

\subsection{RQ2: Influence of the Internationality of Scratch}

\subsubsection{Dataset}

To investigate the influence of the natural language for RQ2, we made
use of two datasets. Our first dataset is a subset of the sprites used
for the evaluation of RQ1. This subset aims to simulate a setup
similar to the method-naming machine learning task when applied to
regular text-based programming languages.
We only retained sprites the name of which exclusively contains
spaces, digits, and characters of the core Latin alphabet~(\ie a to z
in lower- and uppercase, excluding letters with diacritic marks like
ä, and other letter variants like æ). This filtering removed
approximately \perc{5} of the sprites for each of the three dataset
splits.
We applied the same sprite name normalisation steps as for the full
sprite naming dataset of RQ1.

To allow for a more fine-granular analysis, we created a second
dataset by crawling additional projects not already part of the
training dataset of RQ1. We required the projects to contain at least
one sprite that contains more than 20 user-defined
strings~(\ie~variable names or string constants).
Based on those user-defined strings and the sprite name we then
employed the \properNoun{lingua-rs}~\cite{Stahl2023lingua} tool to
determine the natural language used by the programmer. The tool
supports the detection of most languages also supported by the
\Scratch user
interface.\footnote{\url{https://github.com/pemistahl/lingua-rs/tree/v1.7.2?tab=readme-ov-file\#3-which-languages-are-supported},
accessed 2026-01-15}
We only considered sprites where \properNoun{lingua-rs} reported at
least \perc{90} confidence in detecting the correct language and
excluded the stage and sprites using the default
name~(cf.~\cref{sec:setup:sprite-naming-dataset}). Finally, the sprite
name normalisation was applied as for our previously described
datasets.
We limit our evaluation to the ten most frequently occurring languages
for which we could randomly sample \numSpritesPerLanguage sprites
matching these criteria. While six of these languages are using the
Latin alphabet, also Russian written in Cyrillic script, and the
Chinese, Japanese, and Korean writing systems are represented in the
resulting dataset.

\subsubsection{Methodology}

We evaluated all models used in RQ1 again on the two new datasets.
When prompting the \LLMs, we used the same prompt template as for
RQ1~(cf.~\cref{fig:sprite-naming-prompt}).
Since the Latin-only subset of the sprite naming dataset is large
enough to train the non-\llm models specifically for this dataset,
trained all models accordingly and report the performance of models
trained on the latin-only subset. We used the same hyperparameters as
in RQ1 to train the models.
Since the language-specific datasets are too small for dedicated
training, we used the models pre-trained for RQ1 and evaluated them
without further fine-tuning.
We limited our evaluation metrics  on the language-specific datasets
to the F1, \bleu, and \rouge-2
scores~(cf.~\cref{sec:rq1:methods-eval}). Since the computation of the
\meteor metric assumes an English vocabulary, we excluded it for this
experiment.

\subsubsection{Statistical Analysis}\label{sec:rq2-setup:stats}

Following the recommendations of
\citeauthor{Demsar2006Statistical}~\cite{Demsar2006Statistical}, we
used the Friedman test~\cite{Friedman1937Use} with a significance
level of \(\alpha = 0.05\) to compare the models on the ten
language-specific datasets.
A Shapiro-Wilk test~\cite{Shapiro1965analysis} showed that some per-model
scores are non-normal. Thus, we report the median~(\(\MD\)) and
interquartile range~(\(\IQR\)) rather than mean values when comparing
the results of different models.
We used the Nemenyi post-hoc test~\cite{Nemenyi1963Distribution} to
declare any pair of models whose mean rank difference exceeds the
critical distance~(\(\CD\)) as significantly different.
To compute the effect size of differences, we used Vargha-Delaney’s
\(\hat{A}\)~\cite{Vargha2000Critique}.

\subsection{RQ3: Fine-Tuning Sprite Embeddings for Program
Classification}

\subsubsection{Datasets}

Instead of labelling the programs per-sprite, the datasets used to
answer this research question assign one or more labels to the overall
program.

\begin{figure}[t]
  \hfill \subcaptionbox{Project Category\label{fig:dataset:category}}{\includegraphics[width=0.45\textwidth]{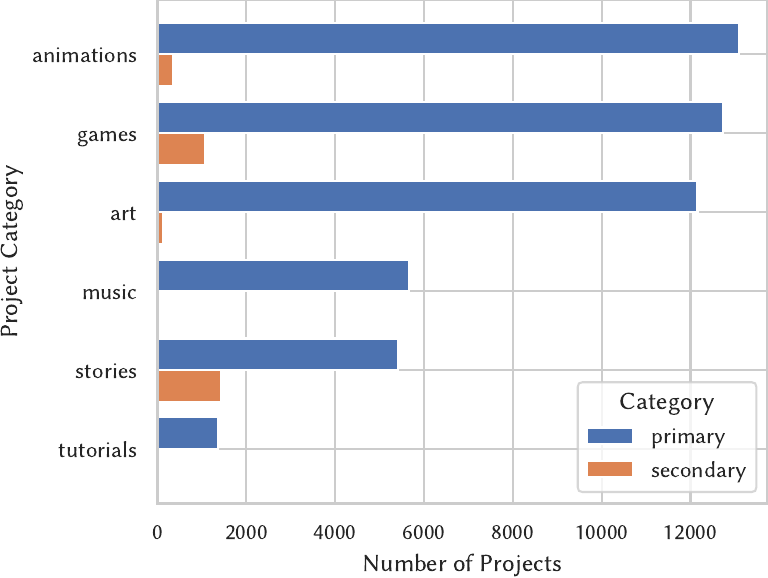}
  }
  \hfill \subcaptionbox{Project Remixes\label{fig:dataset:remixes}}{\includegraphics[width=0.45\textwidth]{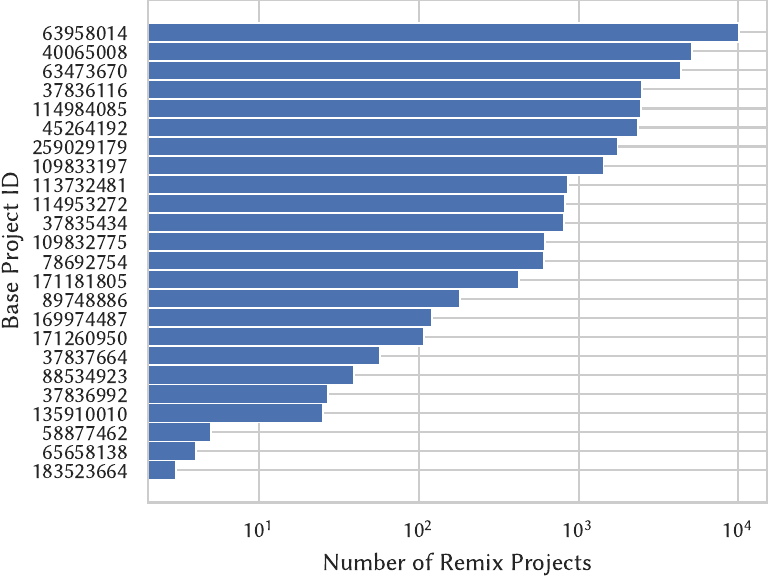}
  }
  \hfill \caption{Project classification dataset label distributions.}
\end{figure}
 
\paragraph{Project Category}

While many \Scratch projects represent interactive games, there also
exist projects that instead focus on presenting a story shown as
non-interactive animations. Other projects showcase visual or musical
arts, for example using the \enquote{Pen} or \enquote{Music}
extensions provided by the \Scratch website.
To evaluate the models’ ability to classify whole projects according
to such categories, we used a dataset of prior
research~\cite{Fein2022Evaluation} after excluding all projects that
were already part of the training dataset of our RQ1 from the
validation and test splits of this dataset.
The dataset was created by labelling \exnum{50560} projects according
to six different base categories~(games, animations, art, music,
stories, tutorials). Projects of the animations, games, and art
categories are the most common ones with similar numbers of projects
per category~(cf.~\cref{fig:dataset:category}). Since programs can fit
into multiple categories, some of them have been assigned up to three
labels. However, most projects~(\exnum{47361}) are labelled only with
a single primary category.

\paragraph{Project Remixes}\label{sec:methods:project-remix-dataset}

As a second dataset, we used \exnum{24} starter projects provided by
the \properNoun{RaspberryPi
Foundation}{}\footnote{\url{https://projects.raspberrypi.org/en/collections/scratch},
\Scratch user CodeClubRik
\url{https://scratch.mit.edu/users/CodeClubRik/}, accessed 2026-06-05}
which are specifically designed to be used as a starting point for own
projects when learning to program in \Scratch. Each starting project
is accompanied by a tutorial explaining steps to implement additional
functionality. However, these tutorials also encourage users to be
creative and extend the projects with additional or different
features. \Scratch users frequently share such \enquote{remixes}
publicly~(cf.\ \cref{sec:setup:dataset-collection}).
Using the \Scratch-\api which links remixes to the base project, we
collected a total of \exnum{34856} remixes of the \exnum{24} starter
projects~(median \(\MD = 617\) remixes per project). As
\cref{fig:dataset:remixes} shows, some base projects are more popular
than others, resulting in dataset imbalance. The most popular base
project is \enquote{Boat
Race}.\footnote{\url{https://scratch.mit.edu/projects/63958014},
accessed 2026-07-16}
During collection, we excluded projects containing identical code to
the respective starter project. We again ensured that no projects are
contained in both the training dataset split of RQ1 and the validation
and test splits of this dataset.

\subsubsection{Methodology}\label{sec:setup:tuning-method}

We used the \enquote{Project Category} dataset for a multi-label
classification task since each project has been assigned one or more
category labels.
For the \enquote{Project Remixes}, each remix was labelled by the
starter project it was derived from to construct a multi-class
classification task with a single label per project.
In both cases, we use common machine learning metrics like the
accuracy, precision, recall, and F1 score to quantify the prediction
performance. Due to the dataset imbalance, we consider both the
macro-average and class-weighted F1 scores.
Since the project remixes of the second dataset should still have a
somewhat similar structure to the starter project, we assume that the
project embedding vectors as generated by the model should form
clusters around the embeddings of the starter project. To evaluate
whether clusters are formed in embedding space, we computed the
Calinski-Harabasz Index~\cite{Calinski1974dendrite}.

To limit the required implementation and computational effort for the
hyperparameter-tuning of fine-tuning the models used for different
approaches, we only used the best-performing dedicated model of the
previous research questions~(\ggnn) and also the
\qwenembedding~\cite{Zhang2025Qwen3} \llm for the evaluation of this
research question.
As baseline evaluation, we specifically trained \ggnn for the two
project classification tasks. For the further evaluation aiming to
find a suitable fine-tuning approach, we used \ggnn as pre-trained for
the sprite naming classification and \qwenembedding in its
downloadable pre-trained state. To fine-tune \ggnn, it was first
converted into an embedding generation model by removing its decoder
layers. By aggregating the individual sprite embeddings, we could
construct an embedding for the whole program. Unlike for the sprite
naming task, we included the stage and sprites independent of their
name~(\ie even when they had a default name).
\qwenembedding can be prompted with the sprite code in \scratchblocks
format~(cf.~\cref{sec:code-processing:llms}) to directly obtain the
embedding on which a decoder can be trained. For \qwenembedding, we
derived two main variants: The first variant,
\enquote{\qwen-per-sprite}, is similar to our \ggnn approach
and receives the \scratchblocks input per sprite to generate sprite
embeddings which can be aggregated. The second variant,
\enquote{\qwen-whole}, receives the code of all sprites together as a
single input to directly generate a program embedding.

We constructed several approaches with which these base models can be
fine-tuned for the classification task. For each approach, we tested
various hyperparameters~(113 configurations per classification task,
of those 80 for \ggnn and 33 for \qwenembedding). We chose the
configurations performing best according to their F1 score as the
overall best ones.

\paragraph{Dedicated Classification Decoder}

This approach computes the individual sprite-embeddings of the project
and applies a max-pooling to obtain the project representation. This
overall project representation is then fed as input into a small
trainable dedicated network that learns to predict the overall project
label.
We used a Multi-layer Perceptron~(\mlp) and a Support Vector Machine
as implemented by the \properNoun{scikit-learn}
library\footnote{\url{https://scikit-learn.org/1.8/index.html}} as two
alternative approaches to implement this prediction network.
In case of the \mlp, we tried configurations without hidden layers
that map directly from model embedding size to the number of output
neurons~(\eg configuration \([128, 6]\) in case of \ggnn for the
project category task) up to configurations that contain hidden layers
with sizes following decrementing powers of 2~(\eg~\([128, 64, \dots,
8, 6]\)).

\paragraph{Training and Fine-Tuning: \mlp-Decoder}

Instead of using the sprite embedding generator as a fixed model with
frozen model weights, we additionally fine-tuned the encoder part of
the overall new model. To fine-tune the model, we replaced the
original decoder of the model with max-pooling and the \mlp as
described above. We then either trained the full model without
freezing any layers or alternatively kept the encoding layers frozen
for one training epoch to force the decoding layers to adapt to the
encoder without causing the encoder to unlearn prior information.
After this epoch of fine-tuning with the original learning
rate~(\exnum{0.002}), we reduced the learning rate by a factor of 10
and unfreeze the encoder weights to slowly fine-tune the whole model
for the specific task.
Finally, instead of fine-tuning the base model, we also used the
alternative model structure with the replaced decoder to train new
task-specific models.
Due to the required computational capacity of tuning \qwenembedding,
we applied this fine-tuning approach only to the \ggnn base model.

\paragraph{Training and Fine-Tuning: Sequential Sprite Embedding
Aggregation}

In the \Scratch user interface, the user can explicitly sort the
sprites into their preferred order. This order is then also reflected
in the project \json~(cf.~\cref{sec:code-processing:scratch}). The
max-pooling step used by the fine-tuning approaches described above
dismisses this information.
To include it in the model, this approach instead applies recurrent
bidirectional long short-term memory~(\lstm) model
layers~\cite{Hochreiter1997Long} to aggregate the individual sprite
embeddings into the overall project embedding before feeding it into a
fully-connected layer mapping the embedding into the predicted labels.
Since the number of sprites per project is usually
small~(cf.~\cref{sec:setup:sprite-naming-dataset,fig:sprite-naming-dataset-stats}),
an \lstm should be able to sufficiently retain the sequential
information.
During hyperparameter tuning we explored configurations having between
one and four \lstm layers. Optional dropout layers in between had a
dropout probability of up to \perc{50}.
Again, we used this model structure both for fine-tuning and dedicated
training of \ggnn model variants but did not fine-tune the
\qwenembedding \llm.

We dismissed using a convolutional neural network~(\cnn) as an
alternative for the \lstm. Due to the typically low number of sprites
per project, the matrices obtained by stacking the individual sprite
embeddings would be very narrow in one dimension and thus often
require padding to be able to apply the convolution. This would likely
reduce the expressiveness of the resulting project embedding in these
cases.

\subsection{RQ4: Program Embeddings as Surrogate Model for Program
Correctness}

\subsubsection{Dataset}\label{sec:setup:correctness-dataset}

\begin{figure}[t]
  \centering \includegraphics[width=0.5\linewidth]{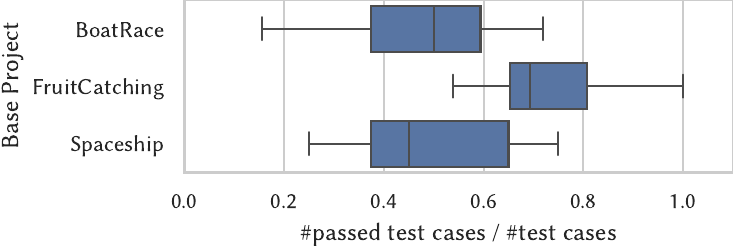}
  \caption{\label{fig:buggy-projects:test-fitness}\Whisker test results of the student projects.}
\end{figure}
 
To answer this research question, we used a dataset from prior
work~\cite{Schweikl2025RePurr} containing three \Scratch projects from
a classroom
setting~(Boat\-Race,\footnote{\url{https://scratch.mit.edu/projects/63957956/},
accessed 2026-05-20} Fruit\-Catching, and Spaceship). For each of the
projects there is at least one bug-free model solution and multiple
incomplete or faulty student solution attempts. Each base project is
accompanied by a \Whisker~\cite{Deiner2023Automated} test suite that
verifies whether the required functionality is implemented correctly
in the student projects. Most students implemented a meaningful
proportion of the
functionality~(cf.~\cref{fig:buggy-projects:test-fitness}).

\subsubsection{Methodology}\label{sec:setup:correctness-method}

Using the \Whisker test suites, we define the test
fitness \(f_\text{test}\) as measurement of the correctness of a
project \(p\). This measurement is given by the proportion of passed
test cases in the test suite:
\begin{align}
f_{\text{test}}(p) = \frac{\left|\text{passed test cases for project
  } p \right|}{\left|\text{all test cases of the test suite}\right|}
\end{align}

Alternatively, given a set of student projects \(P\) for the same
task~(\eg BoatRace) we estimate the project correctness for a student
project \(p \in P\) using the model embedding \(\embed(p) \in
\mathbb{R}^n\) and the embedding \(\embed(s)\) of the corresponding
model solution project \(s\) as
\begin{align}
f_{\text{embed}}(p) = 1 - \frac{\lVert \embed(s) - \embed(p)
  \rVert}{\max_{q \in P}\lVert \embed(s) - \embed(q) \rVert}
\end{align}
to obtain the embedding fitness \(f_\text{embed}\) that measures how
far away a project \(p\) is from the model solution \(s\). The
Euclidean distance between \(s\) and \(p\) in embedding space is
normalised to be in the same interval \([0, 1]\) as the test fitness.
By subtracting this distance from \(1\), a fitness of \(1\) represents
a program equivalent to the solution for both \(f_{\text{test}}\) and
\(f_\text{embed}\). The program out of the dataset \(P\) furthest away
from the solution receives an embedding fitness of \(0\).

To obtain the model embeddings \(\embed(p)\), we used
\ggnn~(\(n=128\)) as the best model of our evaluation of RQ1 and RQ2.
As a comparison to large pre-trained \LLMs, we also computed
\(\embed(p)\) with \(n=4096\) using
\qwenembedding~\cite{Zhang2025Qwen3}.
Since the dataset of 41 student projects per model solution is too
small to fine-tune the models to create task-specific whole-project
embeddings, we used the same approach as in RQ3 and computed
individual sprite embeddings using the \ggnn~(pre-trained on the
sprite naming task) and \qwen-per-sprite models and aggregated the
sprite embeddings into project embeddings using
max-pooling~(cf.~\cref{sec:setup:tuning-method}). We again also used
the \qwen-whole variant of \qwenembedding which receives the whole
program as single input to directly generate a program embedding.

Under the assumption that the model embeddings can be used as
surrogate models to estimate program correctness, we should be able to
observe a direct linear correlation between the actual correctness
\(f_\text{test}\) and the embedding distance \(f_\text{embed}\). To
verify this, we computed the linear least-squares regression and
Pearson’s \(r\) for these two sets of measurements.

\subsection{RQ5: Program Embeddings as Student Progress Indicator}

\subsubsection{Dataset}

We used \ScratchLog~\cite{Caspari2023ScratchLog} to track the changes
students made to their \Scratch projects during an exercise class
session~(45\,minutes including an introductory explanation). In this
exercise, the students started from the basic BoatRace project
solution~(cf.\ \cref{sec:setup:correctness-dataset}) and should
implement four additional features that change the gameplay. They were
tasked with adding a rotating wooden plank the boat can crash against.
The game should also be extended such that a randomly placed crab can
be collected to obtain bonus points which are tracked in a newly
introduced variable. On the game ending screen, the plank should be
hidden and the crab should either also be hidden or it should
congratulate the player when they reached at least five bonus points.
These features thus are non-trivial and require more complex
programming constructs such as loops, conditionals, and variables. We
could verify the successful implementation of the requirements using a
\Whisker~\cite{Deiner2023Automated} test suite containing 11 test
cases.
Saving the state of the projects in 1\,minute intervals, we collected
a total of \exnum{1088} projects across 48 students.

\subsubsection{Methodology}\label{sec:setup:rq5-methods}

To track the progress of students in a course using code embeddings,
\citeauthor{Paassen2021Mapping} proposed the
progress-variance-projection~\cite{Paassen2021Mapping}. Given the
embeddings of the starter program and a model solution, it can map the
student programs into two-dimensional space between these two states.
By construction of the projection, the starting program is represented
by the coordinates \((0,0)\) and the solution by \((1,0)\). Thus, the
x-coordinate of a student program’s projection represents the current
progress. The y-axis represents the variance within the student
program population. I.e.\ if two student programs successfully
implemented the same two features with a different approach, the
projection should have an approximately identical x-coordinate but
different y-coordinates.
While the projection was originally used for code embeddings of Python
programs~\cite{Paassen2021Mapping}, its computation is independent of
the underlying embedding generator. Thus, we again used \ggnn and the
two \qwenembedding variants as in RQ4~(cf.\
\cref{sec:setup:correctness-method}) as pre-trained models without
further fine-tuning to generate the \Scratch project embeddings.

Assuming the resulting projection still reflects the student progress
in our \Scratch setting like it did in the original projection
evaluation on Python code~\cite{Paassen2021Mapping}, even though we
now also use other embedding models, we should be able to observe a
direct linear correlation between the test fitness \(f_\text{test}\)
as defined in RQ4 and the projected student progress on the x-axis. In
both cases values close to \(1\) represent programs close to the
solution. We verify whether this expected correlation exists using a
linear least-squares regression and Pearson’s \(r\).

\subsubsection{Statistical Analysis}

Repeating the linear least-squares regression separately for the
programs of each student, we obtained a population of \(r^2\)
coefficients we used for a statistical comparison between the models.
We applied the same statistical tests as described in
\cref{sec:rq2-setup:stats}.

\subsection{Threats to Validity}

\paragraph{External}

Threats to external validity arise from the sampling of projects for
the model training datasets. Since the \Scratch community tends to
create programs reflecting recent events or popular
media~\cite{Grassl2022Scratch}, existing datasets
(\eg~\cite{Aivaloglou2017Dataset}) as well as ours may be biased
towards topics trending in the community during the dataset crawling.
We mitigate this by sampling projects created over multiple
years~(between May 2021 and April 2023).
A further threat to external validity may be the choice of machine
learning models. We partially mitigate this by using a variety of
model designs using different levels of abstraction from the program’s
\AST to estimate the influence of structural program information and
by comparing to multiple open or proprietary \llm{}s.
A threat to external validity remains, since all models used in our
experiment assume a text- or \AST-based input rather than using visual
code information. The design of the \Scratch workspace results in
multiple related challenges for the application of visual models~(\eg\
scaling of the code area, overlapping blocks) which warrant dedicated
future research on the applicability of such models to \Scratch.
The choice of sprite naming as an evaluation task highlights another
threat to external validity, since by itself it is of limited
relevance to the educational community. However, to train the machine
learning models a suitably large dataset is required, which is not
available for most educational tasks due to the usually small
classroom sizes. We therefore use sprite naming as a pre-training task
for which we can estimate the initial model performance, before
applying the pre-trained models to education-focussed tasks (RQ4 and
RQ5).

\paragraph{Internal}

A threat to internal validity may arise due to bugs in the model
implementations. We try to mitigate this by re-using the
implementations from prior work~\cite{Fein2026Challenges}. To adapt
the models for our \Scratch evaluations, only the preprocessing steps
have been changed. By implementing the preprocessing as part of a
shared tool for all models, we ensure a fair comparison between
models.

\paragraph{Construct}

We mitigate a threat to construct validity by using existing model
implementations used in prior research~\cite{Fein2026Challenges}, thus
reducing the risk of introducing additional bugs.
Since the \Scratch environment allows arbitrary text as identifiers
for the sprites rather than the limited set of alphanumeric characters
allowed in most programming languages~(cf.\
\cref{sec:setup:sprite-naming-dataset}), the subtokenisation
approaches of the existing model approaches might not be optimal for
\Scratch. Instead, subword tokenisation using byte-pair
encodings~\cite{Sennrich2015Neural} or the SentencePiece
tokeniser~\cite{Kudo2018SentencePiece} might achieve better results.
By choosing the existing model implementations we prioritise construct
validity and accept this possible threat to external validity, since
this paper aims to provide a first overarching evaluation of the
applicability of code embeddings for \Scratch rather than optimising
any specific approach. Most tasks in our evaluation~(RQ3--RQ5) also
use the model embedding directly and are thus more independent of the
chosen token en- and decoder approach.

\section{Results}\label{sec:results}

\subsection{RQ1: Using Code Embeddings to Name
Sprites}\label{sec:results-rq1}

\begin{table}[t]
  \caption{Model performance for the sprite naming task on the full dataset and on the dataset containing only names consisting of characters from the Latin alphabet. Best scores per dataset highlighted in bold.\label{tab:sprite-naming}}
  \small \sisetup{round-mode=places}
\begin{tabular}{@{}*{2}{l}*{5}{S[table-format=1.3]}@{}}
\toprule Dataset
& Model
& \multicolumn{1}{c}{F1}
& \multicolumn{1}{c}{top-10 acc.}
& \multicolumn{1}{c}{\textsc{bleu}}
& \multicolumn{1}{c}{\textsc{rouge}-2}
& \multicolumn{1}{c}{\textsc{meteor}}\\
\midrule Full
    & ASTNN
    & 0.3418468285671308
    & 0.4795743322020808
    & 0.5673127551051049
    & 0.1156991787224359
    & 0.1705548790011536\\
    & code2seq
    & 0.2587871381224362
    & 0.5173179569096505
    & 0.4032325901081651
    & 0.0930310471296621
    & 0.1451519526935427\\
    & code2vec
    & 0.2824077694469856
    & 0.5240313305056078
    & 0.4159189916365481
    & 0.1054313801447618
    & 0.1509570606855123\\
    & DeepSeek
    & 0.0039271144914958
    & 0.0041505416106141
    & 0.048075550536984
    & 0.0035441014403136
    & 0.0447662959089453\\
    & Gemma
    & 0.0807736759172059
    & 0.1045333469278513
    & 0.0993842689019848
    & 0.0075021153983404
    & 0.0671908563123805\\
    & GGNN
    & \bfseries 0.5500834030924877
    & \bfseries 0.581037178242853
    & \bfseries 0.7829667216412514
    & \bfseries 0.161779827556244
    & \bfseries 0.4696286008468903\\
    & GPT
    & 0.0764165594364936
    & 0.086467339020521
    & 0.1405059221830506
    & 0.0230004539547387
    & 0.1099691687180247\\
    & Mistral
    & 0.0151877728067623
    & 0.0204950673106566
    & 0.0407184567333282
    & 0.0029238522737501
    & 0.0347578671035063\\
    & NeuralCodeSum
    & 0.1593625039784583
    & 0.263431219159947
    & 0.2296401474765563
    & 0.0416101882574208
    & 0.1377683057450889\\
\midrule Latin Only
    & ASTNN
    & 0.3441951953629201
    & 0.4937508818571354
    & 0.5827587969868089
    & 0.1288298668600921
    & 0.1743342226399304\\
    & code2seq
    & 0.258717718573488
    & 0.5205100567407618
    & 0.4053827366765011
    & 0.0944612907172284
    & 0.1450610929052249\\
    & code2vec
    & 0.2797260314363735
    & 0.5231588022456197
    & 0.4141435917716753
    & 0.1088078375000874
    & 0.1491061209274522\\
    & DeepSeek
    & 0.0042551008613458
    & 0.0042924847664184
    & 0.050993671749443
    & 0.0037214210421205
    & 0.0474827283268525\\
    & Gemma
    & 0.0936830335119361
    & 0.1150970082086471
    & 0.1204871569870209
    & 0.0095400623106619
    & 0.0816580043631826\\
    & GGNN
    & \bfseries 0.583817480883201
    & \bfseries 0.6174424484055259
    & \bfseries 0.8437653450802702
    & \bfseries 0.2224338833262987
    & \bfseries 0.5457714712325278\\
    & GPT
    & 0.0813854620910311
    & 0.0905092131217396
    & 0.1485097077638797
    & 0.0239112803641182
    & 0.1164239874787239\\
    & Mistral
    & 0.0166925530792021
    & 0.0219961769524849
    & 0.0440036005607487
    & 0.0029090526519731
    & 0.0370639780046688\\
    & NeuralCodeSum
    & 0.1646791676944289
    & 0.2756616137642787
    & 0.2433126260952562
    & 0.0439041044085875
    & 0.1455669068679188\\
\bottomrule \end{tabular} \end{table}
 
When naming sprites based on the code within on the full dataset
containing Unicode sprite names, the \ggnn model works
best~(\metricValue{F1}{0.550}, cf.\ \cref{tab:sprite-naming}). Our
results show a clear gap to the next best model
\astnn~(\metricValue{F1}{0.342}), followed by a similar performance of
the \codeToVec and \codeToSeq models. The best model not using
structural code information is
\neuralcodesum~(\metricValue{F1}{0.159}). None of the \llm{}s used in
our experiment~(\deepseekcoder, \gemma, \properNoun{GPT-5-mini},
\mistral) can reliably suggest the correct sprite
name~(\metricValue[\leq]{F1}{0.081}, \metricValue[\leq]{top-10
accuracy}{0.105}).

The low performance of all \llm{}s is somewhat surprising since
\llm{}s have been successfully applied to \Scratch, for example to
offer coding support to
students~\cite{Chen2024ChatScratch,Fein2026Challenges,Fein2025LitterBox+}.
Like our experiments, these other approaches used the \scratchblocks
format or similar ones when prompting the \llm{}s. This suggests that
the \scratchblocks code might be similar enough to regular
code~(cf.~\cref{fig:llm-representation}) to allow the \llm to transfer
learned information into the new domain. However, the coding support
tasks generating natural language descriptions are by design more
open-ended whereas the sprite-naming task requires the generation of a
specific string.
This might be more challenging for the \llm{}s, since unlike regular
code, \Scratch rarely uses custom procedures or variables. Instead,
the most common user-defined textual program elements are string
constants used for program events or exclamations of the animated
sprites. Since the \llm{}s treat the code as a flat token sequence,
these models might rely more on the names of variables and called
methods when processing the code of regular programming languages.
Since this information is rarely available on \Scratch, the structure
of the code becomes more important instead.
Thus, we observe the gaps in performances between models which use a
flat token-based representation of the code~(\llm{}s, \neuralcodesum),
the models using an abstracted representation of the abstract syntax
tree~(\codeToSeq, \codeToSeq, \astnn), and \ggnn which additionally
uses control- and data-flow information.

While the F1-score and top-10 accuracy only take perfectly matching
full names into account, the \bleu, \rouge-2, and \meteor metrics
evaluate how frequently the predicted names are at least partially
correct.
The \ggnn model performs best according to all evaluated
subtoken-based metrics, again followed by \astnn as second-best model.
However, we observe a discrepancy in performance between high \bleu
scores and low \rouge-2 scores for all models~(\eg \ggnn:
\metricValue{\bleu}{0.783}, \metricValue{\rouge-2}{0.162}, \astnn:
\metricValue{\bleu}{0.567}, \metricValue{\rouge-2}{0.116}). The \bleu
score only checks whether correct individual subtokens appear in the
predict name, the \rouge-2 score also checks for matching bigrams of
tokens to evaluate whether they appear in the correct order.
Since \astnn and \codeToVec predict names as a whole instead of
generating them as a sequence of subtokens, the low \rouge-2 scores
might be a result of the models generating an incorrect name which
however generally matches the correct theme and thus shares subtokens
with the original~(\ie resulting in a higher \bleu score).
The sprite names in \Scratch can consist of arbitrary Unicode
characters. Since this makes them potentially more complex than the
method names of text-based languages, the sequence-generating decoders
of the \codeToSeq and \ggnn models might not be sufficient for the new
task. Thus, in case of these models the low \rouge-2 scores might be a
result of the decoder layers generating correct individual subtokens
which are however in the wrong order or interspersed with incorrect
tokens. A more advanced subtokenisation~(\eg
byte-pair-encoding~\cite{Sennrich2015Neural}) rather than the used
natural splitting used in our
implementation~(cf.~\cref{sec:setup:sprite-naming-dataset}) could
improve performance.

Since even the best-performing \ggnn model often suggested incorrect
names and to further investigate the reasons for the \bleu and
\rouge-2 score discrepancies, we manually inspected a subset of 100
random samples where \ggnn predicted an incorrect name.
In \exnum{55} cases the model predicted a name completely distinct
from the original one. For \exnum{40} of those \exnum{55} samples the
original name consists of a single token. In case the prediction was
partially correct, reasons for incorrect predictions are missing
subtokens~(\exnum{4} samples), or the prediction containing one or
more incorrect subtokens~(\exnum{19} samples). However, the most
common failure cause for partially correct predictions are repeated
subtokens~(\exnum{22} samples). Since these repeated subtokens appear
at least once in the original name, its first instance increases the
\bleu score whereas the repetition does not improve the \rouge-2
scores due to non-matching bigrams.
Our findings align with prior work on Java method naming using the
same model implementation~\cite{Fein2026Challenges}, highlighting that
the decoder part of the \ggnn model likely offers an opportunity for
future improvements.

While for some metrics the other model can achieve similar performance
as \ggnn, the model clearly outperforms the others when considering
the \meteor score~(\ggnn: \exnum{0.470}, next best: \astnn with
\metricValue{\meteor}{0.171}). Since the \meteor metric mainly uses
unigrams like \bleu, the repeated subtokens of \ggnn have a limited
impact, especially since the \meteor metric is weighted to favour
recall over precision~\cite{Banerjee2005METEOR}. Since \ggnn also can
frequently produce a valid sequence of multiple subtokens before
starting the repetition, the alignment-penalty of the \meteor metric
likely remains small for many samples.
Since \astnn predicts full names rather than sequences of subtokens,
the alignment-penalty might be higher in case of non-perfect matches.
As indicated by its \bleu score of \exnum{0.403}, the second-best
sequence-generating model, \codeToSeq, likely does not achieve a high
\meteor score~(\exnum{0.145}) since it less often suggests relevant
subtokens.

\begin{summarybox}{1}
Models integrating more structural information about the code
  achieve a better performance. Since \Scratch code rarely uses
  user-defined identifiers, models using flat code representations
  may have more difficulty inferring the code’s intention.
Further improvements can likely be achieved by more advanced name
  subtokenisation or decoder approaches.
\end{summarybox}

\subsection{RQ2: Importance of Natural Language for the Sprite Naming
Task}

\begin{table}[t]
  \caption{Model performance for projects using different natural language strings.
    Best scores per language highlighted in bold.\label{tab:per-language}
  }
  \setlength{\tabcolsep}{2pt}
  \small \resizebox{\textwidth}{!}{\sisetup{round-mode=places}
\begin{tabular}{@{}*{2}{l}*{ 9 }{S[table-format=1.3]}@{}}
\toprule Language
& Metric
& { ASTNN }
& { code2seq }
& { code2vec }
& { DeepSeek }
& { Gemma }
& { GGNN }
& { GPT }
& { Mistral }
& { NeuralCodeSum }\\
\midrule Chinese
    & F1
    & 0.1949125424298867
    & \bfseries 0.2069852016660132
    & 0.192311775352468
    & 0.0037915105029198
    & 0.0146946567321601
    & 0.1549021769543325
    & 0.0694005886985066
    & 0.0114717086567211
    & 0.170668874749579\\
    & \bleu
    & \bfseries 0.2741591729380165
    & 0.2210378585899675
    & 0.1991191140413196
    & 0.0062716720434603
    & 0.0219644719263422
    & 0.1784708603937597
    & 0.0784767054450244
    & 0.0146028397565115
    & 0.1694600812686706\\
    & \rouge-2
    & \bfseries 0.0115065371236655
    & 0.0019414962464405
    & 0.0012913223140495
    & {{\(<\exnum{e-3}\)}}
    & {{\(<\exnum{e-3}\)}}
    & 0.0103581929965249
    & 0.0010505625593059
    & {{\(<\exnum{e-3}\)}}
    & {{\(<\exnum{e-3}\)}}\\
\midrule English
    & F1
    & 0.2498462319056677
    & 0.1467218148137312
    & 0.1169073011853113
    & 0.0042735042735042
    & 0.0637955096478569
    & \bfseries 0.3701194004256732
    & 0.0785950510807492
    & 0.0125944970608211
    & 0.051424046953501\\
    & \bleu
    & 0.4553944280577268
    & 0.3084995598618376
    & 0.2240965162825123
    & 0.0712404969595461
    & 0.1398315931500227
    & \bfseries 0.6585794242496132
    & 0.20841240868889
    & 0.0618278689222038
    & 0.106052347191456\\
    & \rouge-2
    & 0.1020118756057443
    & 0.0830536841659251
    & 0.0760701283307328
    & 0.007385039203221
    & 0.0178765255729049
    & \bfseries 0.1497522159631257
    & 0.0538181946502928
    & 0.0068057827713494
    & 0.0254213458150171\\
\midrule French
    & F1
    & 0.1432665323858625
    & 0.0734299631401506
    & 0.0468420703736062
    & 0.0017251877410188
    & 0.0379477344342003
    & \bfseries 0.2621088308253146
    & 0.0556035186136649
    & 0.0097416435985374
    & 0.0110277852474683\\
    & \bleu
    & 0.2771992535641353
    & 0.1334694173708824
    & 0.0793669416262467
    & 0.0459402872697097
    & 0.0723540821161244
    & \bfseries 0.3966228944953984
    & 0.1387901846380999
    & 0.0344164830904593
    & 0.0255746166197425\\
    & \rouge-2
    & \bfseries 0.0382870726171114
    & 0.023553313350517
    & 0.0244382964453386
    & 0.0050334889385021
    & 0.0052023844981719
    & 0.0349120086233556
    & 0.0321427766345599
    & 0.0045550775114389
    & 0.0038653366583541\\
\midrule German
    & F1
    & 0.1562203509583346
    & 0.0655138304728203
    & 0.0447709994334252
    & 0.0032480382122142
    & 0.0505355302448104
    & \bfseries 0.2425021024171312
    & 0.0680078278718064
    & 0.0133800484421831
    & 0.0149718689500096\\
    & \bleu
    & 0.2679822948322859
    & 0.123869693709603
    & 0.0621926159298265
    & 0.0435503420034198
    & 0.0988735671794236
    & \bfseries 0.3970212901940018
    & 0.1533532627807816
    & 0.0390186412976055
    & 0.0274345948500835\\
    & \rouge-2
    & 0.0391837595253155
    & 0.0126298496326905
    & 0.0062365591397849
    & 0.0040139258652467
    & 0.0090459205300556
    & \bfseries 0.0673633510955034
    & 0.0296554077106789
    & 0.0028315343053561
    & 0.0029579502894109\\
\midrule Japanese
    & F1
    & \bfseries 0.1944466561048493
    & 0.1164739944759102
    & 0.1147581556083668
    & 0.0027777777777777
    & 0.0145710689444724
    & 0.1869783180128125
    & 0.0489570781426327
    & 0.0040008932866075
    & 0.0386966416627064\\
    & \bleu
    & \bfseries 0.2713553953834502
    & 0.1589084799859411
    & 0.1354833275095228
    & 0.0091121365354332
    & 0.0225112672260838
    & 0.2370919001715533
    & 0.0695596445964486
    & 0.0101724282305765
    & 0.0513498354307177\\
    & \rouge-2
    & 0.0133459516298633
    & 0.0081044656516354
    & 0.0064753682072117
    & {{\(<\exnum{e-3}\)}}
    & 0.0012249897917517
    & \bfseries 0.0186220601045296
    & 0.0042330202803865
    & {{\(<\exnum{e-3}\)}}
    & 0.0034774593598123\\
\midrule Korean
    & F1
    & 0.103131096963599
    & 0.0255486055947637
    & 0.018582958746632
    & 0.0046897171050669
    & 0.0134331238082304
    & \bfseries 0.1244525062276128
    & 0.0321488237558413
    & 0.0027575164561465
    & 0.0058175115897026\\
    & \bleu
    & 0.1610448773611949
    & 0.0497915328534721
    & 0.0274587999373481
    & 0.0182235567924504
    & 0.0214518782779452
    & \bfseries 0.168673088626445
    & 0.0627774007417476
    & 0.0091247483527807
    & 0.0072434559375115\\
    & \rouge-2
    & \bfseries 0.0125755079626578
    & 0.0051039938752073
    & 0.0029314300280397
    & {{\(<\exnum{e-3}\)}}
    & {{\(<\exnum{e-3}\)}}
    & 0.0111965623877969
    & {{\(<\exnum{e-3}\)}}
    & {{\(<\exnum{e-3}\)}}
    & {{\(<\exnum{e-3}\)}}\\
\midrule Portuguese
    & F1
    & 0.2067671120923959
    & 0.108703927801446
    & 0.0757500558299126
    & 0.005419228308496
    & 0.0803125230630321
    & \bfseries 0.2589850930852128
    & 0.1067400071505609
    & 0.0224177630356827
    & 0.0191183846495801\\
    & \bleu
    & 0.3202746556325967
    & 0.1674462753543778
    & 0.1060033873299083
    & 0.0511602056246797
    & 0.1039571435412406
    & \bfseries 0.402073895966653
    & 0.1803409099887326
    & 0.0429694876064665
    & 0.0435093359751764\\
    & \rouge-2
    & 0.0669984163040595
    & 0.0277262524880362
    & 0.0298938105512543
    & 0.0085177119628339
    & 0.0092163868394043
    & \bfseries 0.0753928067042932
    & 0.0453105239385727
    & 0.0043773684680872
    & 0.0140244416350611\\
\midrule Russian
    & F1
    & 0.1103261082404589
    & 0.0771029008981701
    & 0.0556983032243267
    & 0.0050542979120718
    & 0.0128584426876957
    & \bfseries 0.1421609963419633
    & 0.0546485672594966
    & 0.0054471683123368
    & 0.0300796847046161\\
    & \bleu
    & \bfseries 0.2156196341287048
    & 0.0995719048550128
    & 0.0734369344993303
    & 0.0215408517035349
    & 0.0201112315483647
    & 0.1923525673141183
    & 0.0804542427462917
    & 0.0123678882850014
    & 0.0383938572721189\\
    & \rouge-2
    & 0.0138231258276558
    & 0.0035583797628151
    & 0.0016109203442282
    & {{\(<\exnum{e-3}\)}}
    & {{\(<\exnum{e-3}\)}}
    & \bfseries 0.0315961958617144
    & 0.0013506212857914
    & {{\(<\exnum{e-3}\)}}
    & {{\(<\exnum{e-3}\)}}\\
\midrule Spanish
    & F1
    & 0.1898698185109891
    & 0.0709080370233878
    & 0.0519095381503134
    & 0.0034670796575558
    & 0.0568959337992799
    & \bfseries 0.2855483843400068
    & 0.0828666710177305
    & 0.0126701011189193
    & 0.0157674448518896\\
    & \bleu
    & 0.3212354353943117
    & 0.1379983627206116
    & 0.0812436568797133
    & 0.0452372928392451
    & 0.0874527978341126
    & \bfseries 0.454625821782778
    & 0.1634750114056401
    & 0.0338514043766958
    & 0.0289764928240838\\
    & \rouge-2
    & 0.0655772211069141
    & 0.0237154394930622
    & 0.0210392902408111
    & 0.0031687888830745
    & 0.0051786394433834
    & \bfseries 0.0770151279132011
    & 0.0344869424357125
    & 0.0033647128885224
    & 0.0030459625241576\\
\midrule Turkish
    & F1
    & 0.1264714990323315
    & 0.0457487446065627
    & 0.0345253619019924
    & 0.0032001660384132
    & 0.0320966970671863
    & \bfseries 0.1924781729396723
    & 0.053422962296706
    & 0.0059877227403602
    & 0.0120428706822478\\
    & \bleu
    & 0.2632335031894098
    & 0.0921341208952201
    & 0.0434654891126281
    & 0.0344992402862651
    & 0.0450748781407824
    & \bfseries 0.2895991091992916
    & 0.0992795334948788
    & 0.020758223209734
    & 0.0186345256192037\\
    & \rouge-2
    & 0.0550240466959551
    & 0.0142505416981607
    & 0.0075526783447575
    & 0.0056350488499115
    & 0.0037007132283676
    & \bfseries 0.055450147034032
    & 0.031220751840453
    & 0.0029766571553019
    & 0.001201923076923\\
\bottomrule \end{tabular}   }
\end{table}
 
Reducing the model training and evaluation datasets to sprites with
names that only contain characters of the core Latin alphabet shows no
clear difference in model performance compared to the full Unicode
dataset~(cf.\ \cref{tab:sprite-naming}). The \ggnn model still
performs best and also the pattern of increasing model performance
with increasing structural information used in the model remains
unchanged.
While there is a slight improvement for all metrics compared to the
Unicode dataset~(\eg \ggnn \metricValue{F1}{0.550} vs.\ \exnum{0.584}
and \metricValue{\bleu}{0.783} vs.\ \exnum{0.844}), this can be most
likely attributed to the smaller vocabulary size. The subtokenisation
process and model decoder designs~(cf.\ \cref{sec:results-rq1}) likely
remain as the main factors limiting model performance.

Evaluating the models on dedicated datasets per natural language used
by the programmer can give more insights into the impact of the
natural language on model performance.
As \cref{tab:per-language} shows, again \astnn and \ggnn perform best
in nearly all cases. Only for the programs identified as Chinese, the
\codeToSeq model achieves the highest F1 score~(\exnum{0.207}, \astnn:
\exnum{0.195}). Even though the projects of the per-language datasets
were specifically chosen for containing multiple user-defined natural
language strings and identifiers, the \llm{}s can only rarely predict
accurate names.

Using the 10 languages as population, a Friedman
test~\cite{Friedman1937Use} shows there are significant differences
between the metrics of the nine models~(\pValue[<]{e-11} for all three
metrics).
We then apply a post-hoc Nemenyi test~\cite{Nemenyi1963Distribution}
with \(\CD = \num{3.799}\). For all three metrics, the \ggnn, \astnn,
\codeToSeq, \gpt, and \codeToSeq models form the best-performing group
without significant differences in between~(listed by increasing mean
rank, lower rank is better; same ranking across all three metrics).
The other models~(\deepseekcoder, \gemma, \mistral, and
\neuralcodesum) form a second group with mean rank orders varying by
metric.
Since the \ggnn and \astnn models achieve the highest scores in nearly
all cases, and since they are consistently have the best mean ranks,
the following evaluation focusses on these two best-performing models.

Both \astnn and \ggnn perform best on English-language programs, which
is to be expected since it likely is the most common natural language
in the model training dataset.
Nevertheless, both models perform worse than on the general sprite
naming dataset used for RQ1. While the English-only dataset consists
of sprites with fewer blocks than the general dataset~(\(\md=24\),
\(\iqr=43\); cf.~\cref{sec:setup:sprite-naming-dataset}), these blocks
are combined into more complex programs~(cyclomatic complexity:
\(\md=19\), \(\iqr=32\)). This might be a result of our dataset
creation process specifically selecting sprites that contain multiple
user-defined strings as part of string constants, messages, or
variable names. Especially variables are likely used by more advanced
programmers or in more complex programs. The increase in complexity is
also reflected in an increase in the Halstead difficulty
metric~(\(\md=\num{13.726}\), \(\iqr=\num{21.148}\)), which is
computed based on the number of operators and operators used in the
code~\cite{Halstead1979Elements}.

When comparing the F1 scores~(\ie considering perfect name
suggestions) both \astnn and \ggnn perform worst for Korean and
Russian programs. These two languages likely appear as part of the 10
languages sampled for this research question, since their fairly
unique writing systems allow the \properNoun{lingua-rs}
tool~\cite{Stahl2023lingua} to confidently recognise them. However,
the models likely exhibit low performance since there are
comparatively few \Scratch users living in the respective
countries~(Russia: \perc{0.78} of all users, North Korea: \perc{0.11},
South Korea:
\perc{0.03}\footnote{\url{https://scratch.mit.edu/statistics/},
accessed 2026-05-19. The \Scratch website allows users to select their
country of residence rather than using automatic geolocation. The
actual number of users from North Korea is likely lower.}) and thus
the model training dataset only contains few samples in these
languages.

\ggnn performs best on the six languages using a Latin-based writing
system, whereas for \astnn the best-performing English and Portuguese
datasets are followed by Chinese and Japanese.
In case of \ggnn, the Latin-based languages are likely the most
compatible with our subtokenisation process~(cf.\
\cref{sec:setup:sprite-naming-dataset}). For the Chinese, Japanese,
and Korean writing systems these whitespace- and capitalisation-based
subtokenisation rules may often not apply. Subtokens might also be
more diverse due to a larger number of available characters.
Consequently, there may be only few examples of each subtoken in the
training dataset and thus the model cannot produce valid subtoken
sequences. This is highlighted by the difference in \bleu score for
Chinese programs between \ggnn~(\exnum{0.178}) and
\astnn~(\exnum{0.274}). Using a
byte-pair-encoding~\cite{Sennrich2015Neural} or
\properNoun{SentencePiece} tokenizer~\cite{Kudo2018SentencePiece} to
subtokenise the sprite names instead could likely improve the
performance.
\astnn on the other hand uses the whole name as a single token and
might thus require fewer samples in the training dataset to exhibit
the better performance for Chinese and Japanese.

\begin{summarybox}{2}
As expected, the models perform better on more common natural
  languages. A more advanced subtoken encoding should be used to
  improve the performance for languages not using Latin-based writing
  systems.
\end{summarybox}

\subsection{RQ3: Sprite Embeddings for Program Representation}

\subsubsection{Results}

For this research question we not only fine-tune a sprite-embedding
model to evaluate the classification performance, but also investigate
the expressiveness of the underlying code embeddings.
To limit computational effort, we continue the evaluation only with
the best-performing sprite-naming model~(\ggnn) and two
\qwenembedding-based variants.

\paragraph{Program Classification}

\begin{table}[t]
  \caption{Model performance for the program category and remix classification tasks.
    The Calinski–Harabasz Index~(\CHI) is computed on the respective program embeddings.
    Best scores highlighted in bold.\label{tab:classification-results}
  }
  \setlength{\tabcolsep}{2pt}
  \small \resizebox{\textwidth}{!}{\sisetup{round-mode=places}
\begin{tabular}{@{}*{5}{l}*{5}{S[table-format=1.3]}S[table-format=3.3]@{}}
\toprule Task
& Model
& Model Variant
& Aggregation
& Classifier
& \multicolumn{1}{c}{Acc.}
& \multicolumn{1}{c}{Prec.}
& \multicolumn{1}{c}{Rec.}
& \multicolumn{1}{c}{F1}
& \multicolumn{1}{c}{F1 (weighted)}
& \multicolumn{1}{c}{CHI}\\
    \midrule Category
        & GGNN
        & dedicated
        & max pool
        & MLP
        & 0.49812067260138476
        & 0.696263882112321
        & 0.3882032709634624
        & 0.43961612459075977
        & 0.5548086257706083
        & \bfseries 35.77397954575129\\
        & GGNN
        & fine-tuned
        & \lstm
        & MLP
        & 0.5086053412462908
        & 0.6706983364573237
        & 0.41087547611541203
        & 0.4749721038800292
        & 0.5741202881218224
        & 25.23173193094268\\
        & GGNN
        & sprite embedding
        & max pool
        & Random Forest
        & 0.2973293768545994
        & \bfseries 0.7201673999186688
        & 0.25198956802742517
        & 0.35341035751245387
        & 0.4159244430469313
        & 10.346433193123218\\
        & Qwen 3
        & sprite embedding
        & max pool
        & MLP
        & 0.4929667519181586
        & 0.6103771336553296
        & \bfseries 0.5228089093786521
        & \bfseries 0.561666419212132
        & \bfseries 0.6132839654741308
        & 23.14019973604513\\
        & Qwen 3
        & program embedding
        & none
        & MLP
        & \bfseries 0.5171117705242334
        & 0.7133890287836909
        & 0.4596062591864111
        & 0.5383317049648996
        & 0.6125540392780364
        & 13.547520782520026\\
    \midrule Remixes
        & GGNN
        & dedicated
        & max pool
        & MLP
        & 0.9796386578720964
        & 0.844051767654881
        & 0.8304677256109084
        & 0.8366243611129548
        & 0.9795066938766328
        & 61.83752306150129\\
        & GGNN
        & fine-tuned
        & max pool
        & MLP
        & 0.9856610266704904
        & 0.8612026930239031
        & 0.8429570097772964
        & 0.849142643241564
        & 0.9853363427468127
        & 55.74324555604371\\
        & GGNN
        & sprite embedding
        & max pool
        & Random Forest
        & 0.9716088328075709
        & \bfseries 0.9302813158458454
        & \bfseries 0.8993907721743439
        & \bfseries 0.9127283842548721
        & 0.9701101200985204
        & 199.91027910348492\\
        & Qwen 3
        & sprite embedding
        & max pool
        & MLP
        & \bfseries 0.9876116392970326
        & 0.8898456297597346
        & 0.8848306925275267
        & 0.88698563517653
        & \bfseries 0.9859002980761862
        & 148.00567050287032\\
        & Qwen 3
        & program embedding
        & none
        & MLP
        & 0.9862385321100917
        & 0.8875083530634966
        & 0.8767876139652533
        & 0.8809132175633937
        & 0.984333731758917
        & \bfseries 272.56640119418233\\
\bottomrule \end{tabular}   }
\end{table}
 
For the project category classification task, fine-tuning the \ggnn
sprite embedding model by adding an \lstm aggregation layer followed
by a Multi-layer Perceptron~(\mlp) is the best approach. However, as
\cref{tab:classification-results} shows, the \ggnn model which was
specifically trained from scratch for this task performs nearly as
well~(\ggnn tuned: \metricValue{F1}{0.475}, from scratch:
\metricValue{F1}{0.440}). When using the encoder layers of the sprite
naming model as fixed sprite embedding generators without further
fine-tuning, a Random Forest rather than an \mlp performs best, but
still achieves an overall lower performance~(\metricValue{F1}{0.353})
than the other two \ggnn variants.
Surprisingly, even though the \llm{}s were not able to suggest useful
sprite names, in this task both \qwenembedding model variants
outperform all \ggnn variants. When using the whole program text as
single input to the embedding model, it achieves an F1 score of
\exnum{0.538} and using separate per-sprites inputs followed by
max-pooling the sprite embedding even performs slightly
better~(\metricValue{F1}{0.562}). This performance difference is
equalised by considering the class-weighted F1 score~(both variants
\exnum{0.613}). The accuracy of the \qwenembedding-based
classifiers~(\exnum{0.493} and \exnum{0.517}) is similar to that of
the best \ggnn variant~(\exnum{0.509}), but the \qwenembedding-based
model shows a better balance between precision and recall and thus
achieves an overall better F1 score.

We conjecture that the \llm{}-based classifiers work well for this
task since in some program categories the programs are dominated by
specific kinds of blocks. For example, \enquote{animations} frequently
feature blocks relating to the sprite looks but rarely contain blocks
handling key or mouse input events.\@ \enquote{Games} on the other
hand often feature the latter kind of blocks. The information about
the kinds of used blocks is also part of the flat \scratchblocks token
sequence given to the \llm. Thus embedding structural code information
might be less important for this task.

For the project remix classification task again using a max-pooling
sprite embedding aggregation followed by an \mlp classifier works best
for most base-model and tuning approaches. All model variants can
accurately classify which base project a remixed project was derived
from, resulting in minimal differences in weighted F1 scores between
the models~(range \numrange{0.970}{0.986}). Due to the overall high
scores and closeness of the scores, no clear best model can be
determined.
However, the \ggnn-based classifier with fixed sprite embeddings and a
Random Forest classifier seems to perform best also for classes with
fewer samples since it outperforms the other model variants according
to the non-weighted F1 score.
Since the models can correctly classify most remixes, these results
show that all model variants are robust against small changes in the
program code.

For both the program classification and remix tasks, the relatively
simple max-pooling sprite embedding aggregation is the best
configuration. We hypothesise that the more complex \lstm aggregation
does not result in better model performance, since the number of
sprites in the \Scratch programs is generally quite small~(median: 2,
cf.\ \cref{sec:setup:sprite-naming-dataset}), and thus the benefits
from a sequential \lstm-based aggregation are limited. In some cases
the majority of the code might also be placed in few of the sprites
whereas the other sprites act as static background decoration without
own logic.

\paragraph{Embedding Expressiveness}

\begin{figure}[t]
  \centering \hfill \subcaptionbox{\ggnn}{\includegraphics[width=0.45\textwidth]{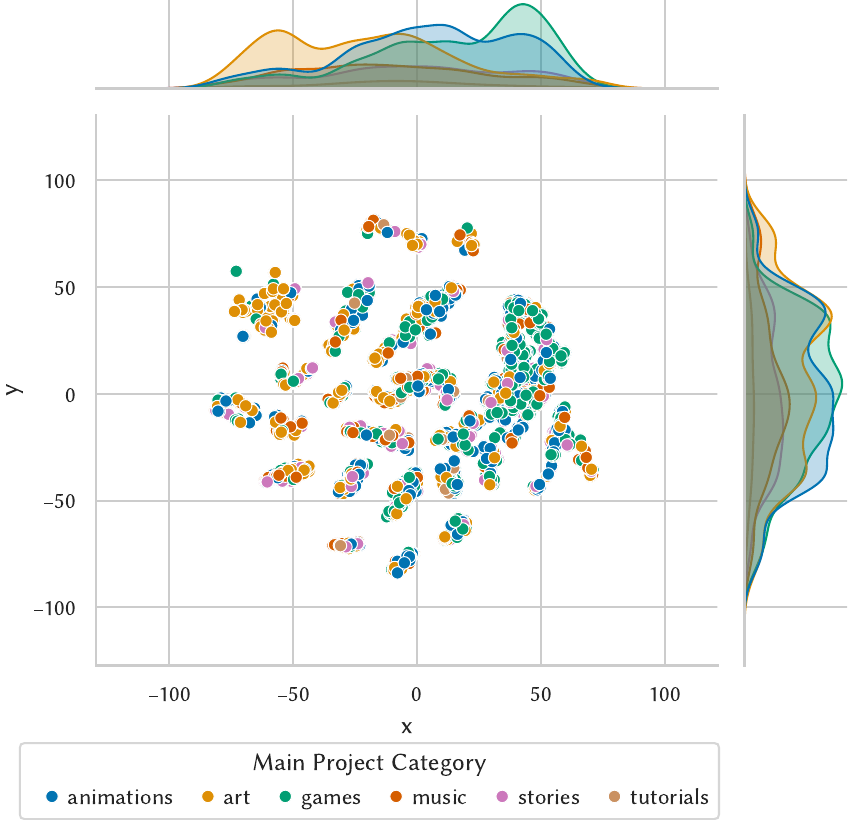}
  }
  \hfill \subcaptionbox{\qwenembedding\label{fig:clustering:category:qwen}}{\includegraphics[width=0.45\textwidth]{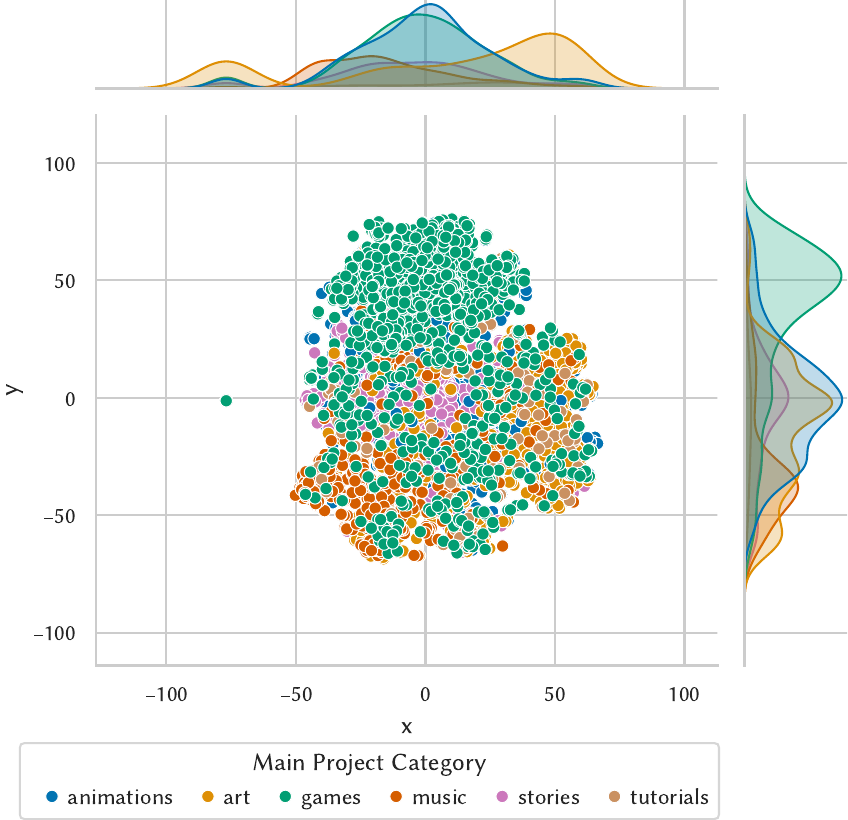}
  }
  \hfill \caption{\label{fig:program-embedding-clustering:category}Project Category:
    Visualisation of the embedding space after t-SNE dimensionality reduction.
  }
\end{figure}
 
Since the classifiers seem to be robust against small changes in the
inputs, their embedding vectors potentially also form clusters of
close vectors for similar inputs.
However, in case of the project category task the embeddings of the
specifically trained \ggnn classifier achieve the highest, \ie best,
Calinski-Harabasz-Index~(\CHI)~\cite{Calinski1974dendrite} even though
the \qwenembedding-based models performed best for the actual
classification task. In case of the project remix task, the
\qwenembedding program embeddings achieve the highest scores followed
by the aggregated but not fine-tuned \ggnn sprite-embeddings.

For the project remix task, the \qwen-based model likely produces
better clusters since the main purpose of such a pre-trained embedding
model is to create similar embeddings for semantically similar inputs.
The remixed projects are likely also often syntactically similar to
the original project.

In case of the project categorisation, the clusters of the
\qwenembedding might not represent the categories relevant for the
task but instead others learned from the training data. For example,
programs relating to characters in the Nintendo universe~(\eg Mario,
Zelda, Donkey Kong) might be clustered due to similar user-defined
strings appearing in the project, even though some of these projects
are games while others are non-interactive animations. The \ggnn
classifier on the other hand is specifically trained on this task, so
its clusters are more likely to align with the labels learned from
the training dataset.

By using a t-SNE projection~\cite{Maaten2008Visualizing} to reduce the
high-dimensional embedding vectors to two dimensions, the model
embedding spaces can be visualised.
\Cref{fig:program-embedding-clustering:category} shows the embedding
projection for the project category task for the---according to the
\CHI---best-performing \ggnn and \qwenembedding model variants. Both
figures show the same number of projects, but in case of \ggnn the
reduction to two dimensions results in many overlapping data points
whereas they are more spread out for \qwenembedding. As the
visualisation of both model embedding spaces shows, this multi-class
classification task might be inherently more difficult for a
classifier model since many categories overlap. For example, it might
be hard to discern whether a program is mainly a game that features
some animations, or whether it primarily focusses on the animation
aspect and contains some interactive elements. This difficulty is
reflected by the low \CHI clustering scores for all model
variants~(cf.~\cref{tab:classification-results}).
Even though the project categories do not form clear clusters, there
seem to be differences in the representations in the project
categories. For example, in case of
\qwen~(cf.~\cref{fig:clustering:category:qwen}), the projects of the
\enquote{animations} and \enquote{games} categories are similarly
distributed on the \(x\)-axis, but clearly different on the
\(y\)-axis.

\begin{figure}[t]
  \centering \hfill \subcaptionbox{\ggnn\label{fig:clustering:remixes:ggnn}}{\includegraphics[width=0.45\textwidth]{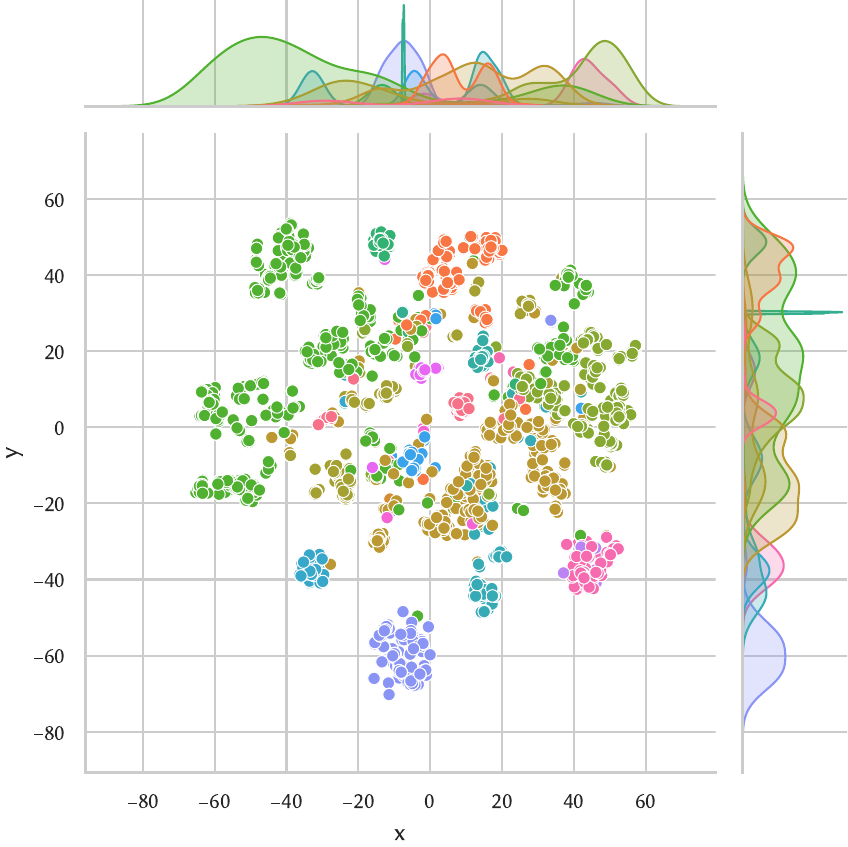}
  }
  \hfill \subcaptionbox{\qwenembedding\label{fig:clustering:remixes:qwen}}{\includegraphics[width=0.45\textwidth]{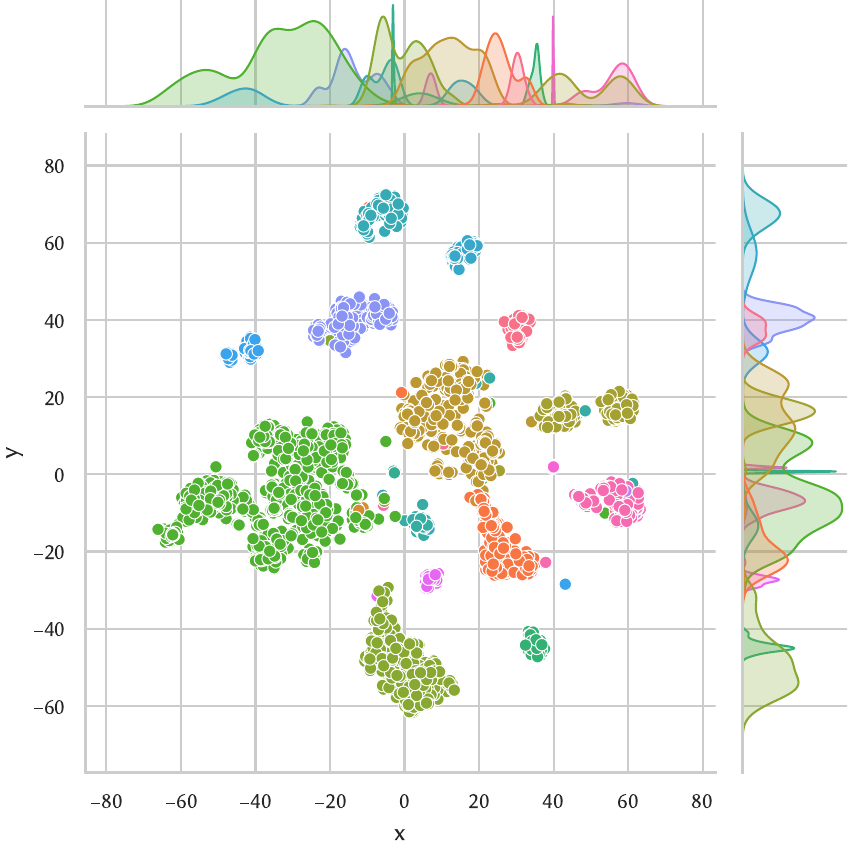}
  }
  \hfill \caption{\label{fig:program-embedding-clustering:remixes}Project remixes:
    Visualisation of the embedding space after t-SNE dimensionality reduction.
    Colouring based on the base project.
  }
\end{figure}
 
For the project remix task on the other hand, the projection of the
\qwenembedding embedding
space~(cf.~\cref{fig:clustering:remixes:qwen}) shows a clear
separation of clusters. For the \ggnn model with a lower
\CHI~(\exnum{199.910} compared to \exnum{272.566} for \qwen), most of
the clusters remain compact, but some are more spread out, \eg~the
green one on the left-hand side~(\(x \approxeq [-70, …, -30]\), \(y
\approxeq [-20, …, 60]\), cf.~\cref{fig:clustering:remixes:ggnn}).
Additionally, the clusters are no longer as clearly separated.
Nevertheless, the visualisations for both models highlight that the
overall classifier models are not only robust against small changes to
the inputs, but also that similar inputs yield similar embedding
vector representations. At the same time, they are sufficiently
expressive to map unrelated code into separate clusters.

\begin{summarybox}{3}
Fine-tuning the pre-trained sprite-embedding models for a new
  classification task can achieve similar performance as task-specific
  models.
Both the overall classifier models and the embedding vectors are
  robust against small changes in the program code. The embeddings can
  be used to form clusters of related programs.
\end{summarybox}

\subsubsection{Discussion}

Our results show that especially approaches combining the pre-trained
sprite embeddings as input for a small task-specific classifier model
allows for a computationally efficient adaption of the models to a new
task. This lowers the required computational effort when designing new
embedding-supported tools.
This confirms that such models are well suited for downstream
educational tasks. For example, an immediately possible educational
application for the clustering of similar projects could be the
recommendation of \Scratch projects based on the learner’s skills or
preferences to encourage them to iteratively explore more challenging
programming concepts~\cite{Qi2022Scratch}.

\subsection{RQ4: Program Embeddings as Surrogate Model for Program
Correctness}

\subsubsection{Results}

\begin{figure}[t]
  \centering \begin{subcaptionblock}{\textwidth}
    \includegraphics[width=0.33\textwidth]{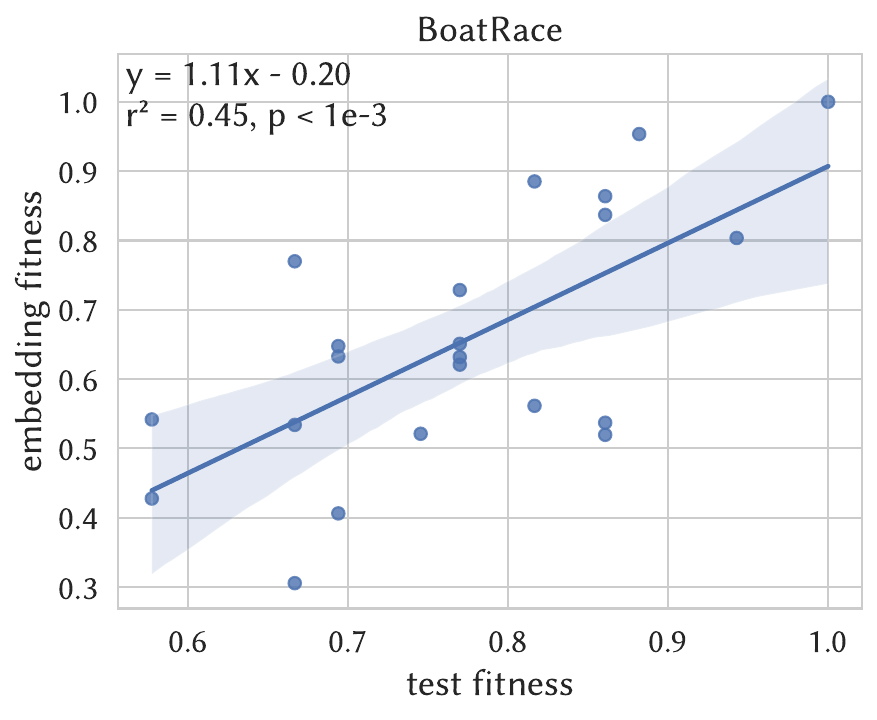}
    \includegraphics[width=0.33\textwidth]{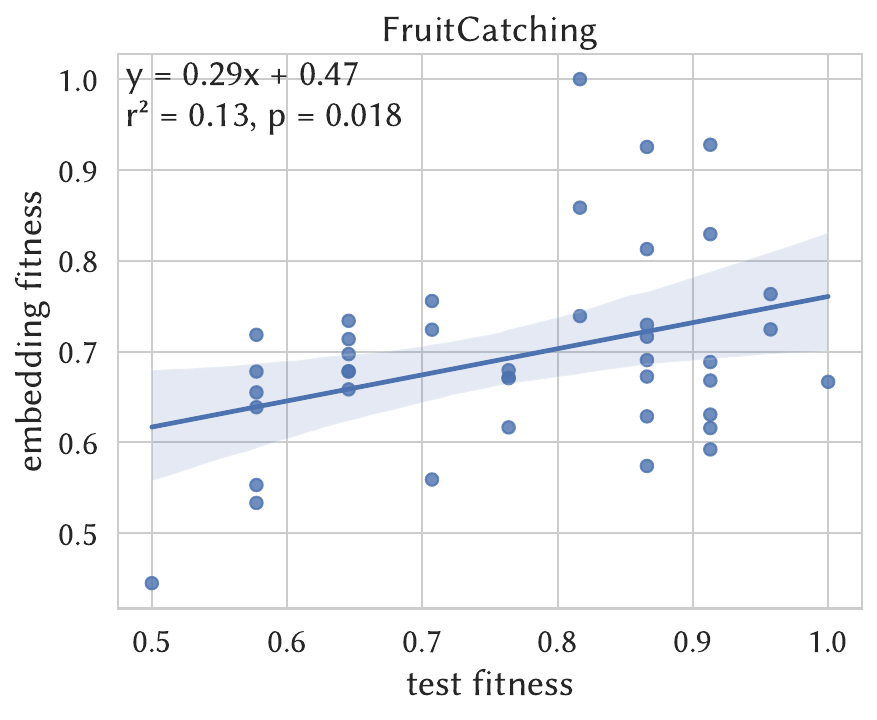}
    \includegraphics[width=0.33\textwidth]{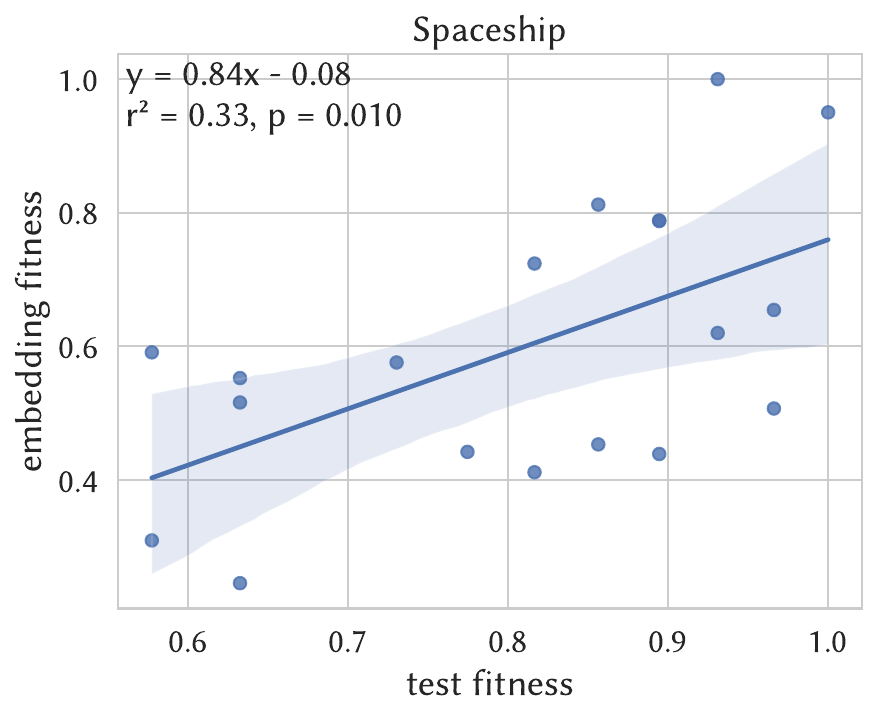}
    \caption{\ggnn per-sprite embeddings followed by max-pooling aggregation.\label{fig:buggy-projects:ggnn}}
  \end{subcaptionblock}
  \begin{subcaptionblock}{\textwidth}
    \includegraphics[width=0.33\textwidth]{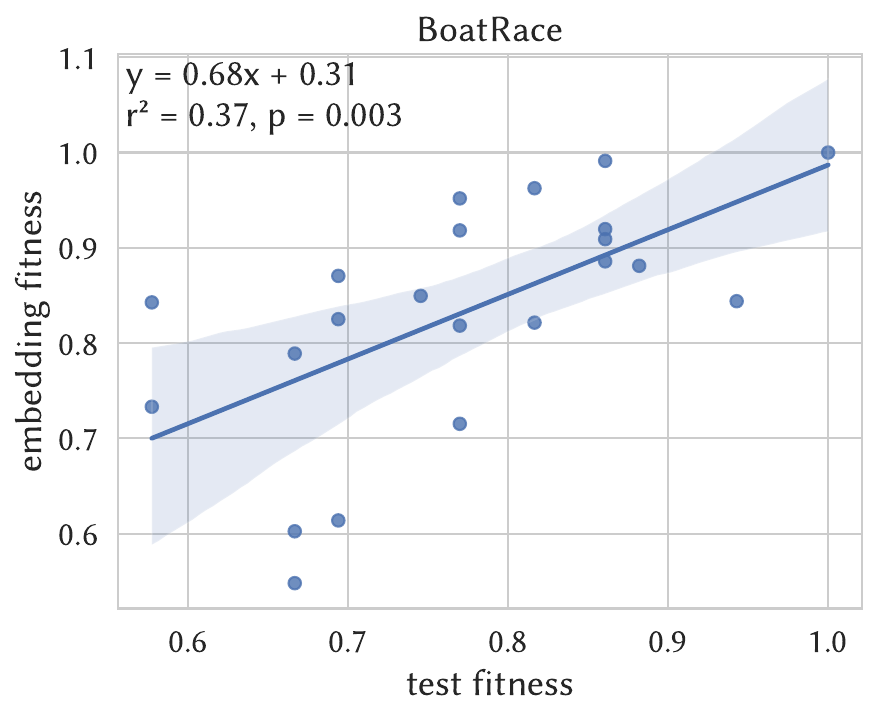}
    \includegraphics[width=0.33\textwidth]{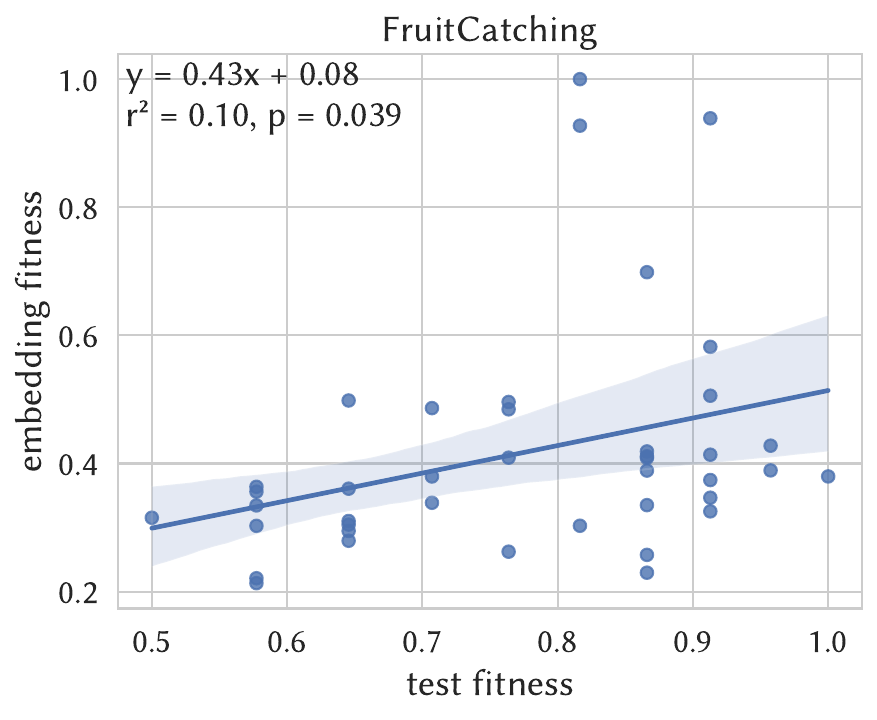}
    \includegraphics[width=0.33\textwidth]{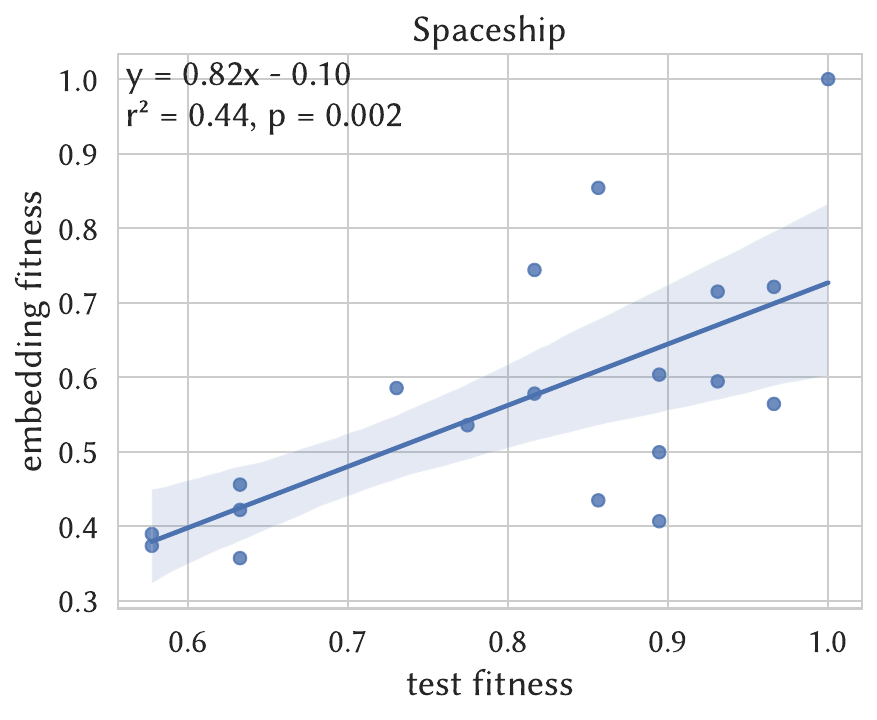}
    \caption{\qwenembedding with the whole projects’ \scratchblocks code as single input.\label{fig:buggy-projects:qwen-whole}}
  \end{subcaptionblock}
  \begin{subcaptionblock}{\textwidth}
    \includegraphics[width=0.33\textwidth]{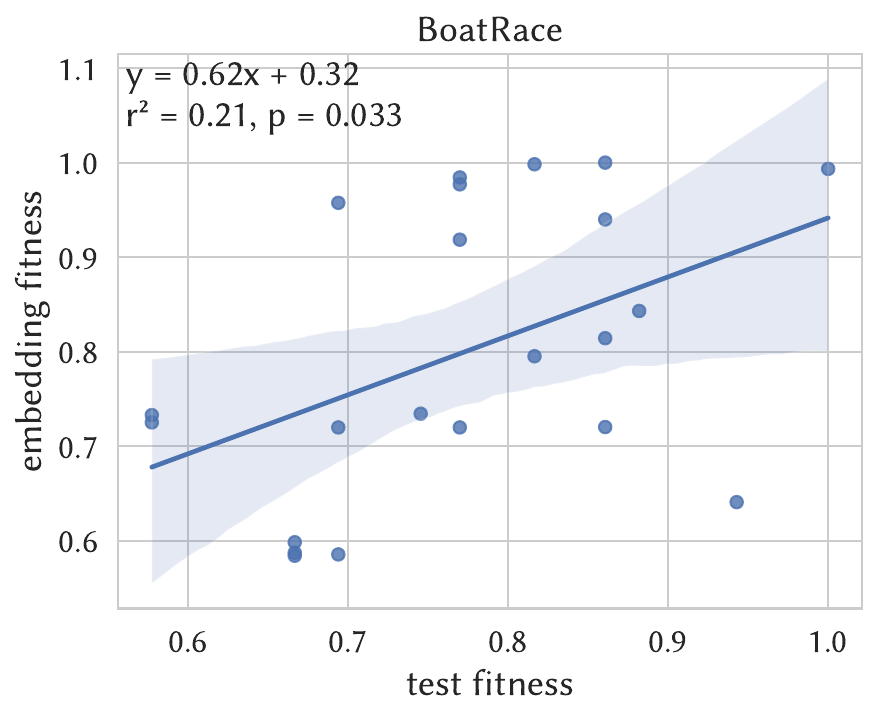}
    \includegraphics[width=0.33\textwidth]{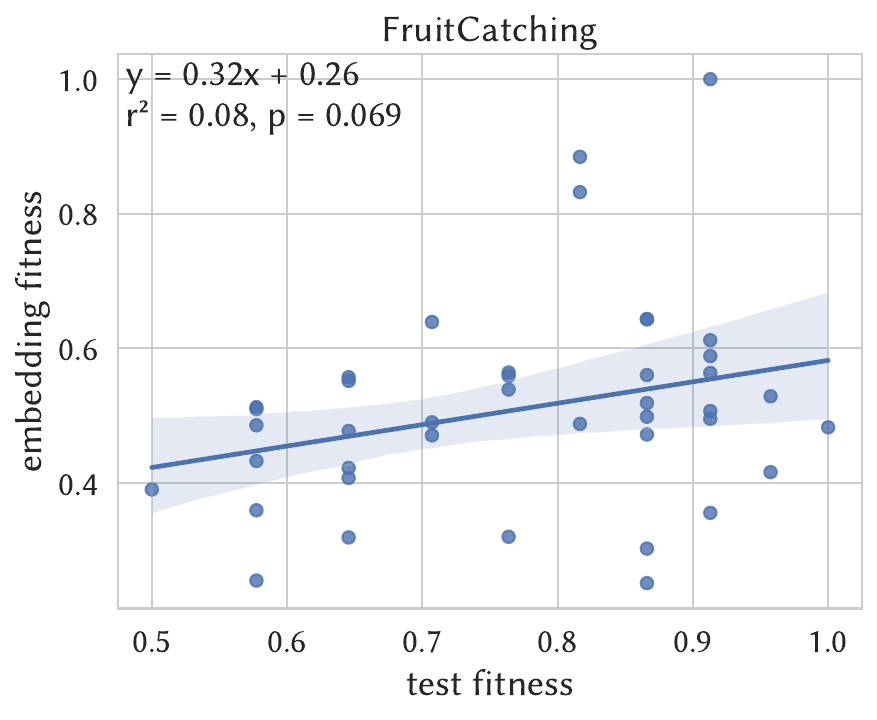}
    \includegraphics[width=0.33\textwidth]{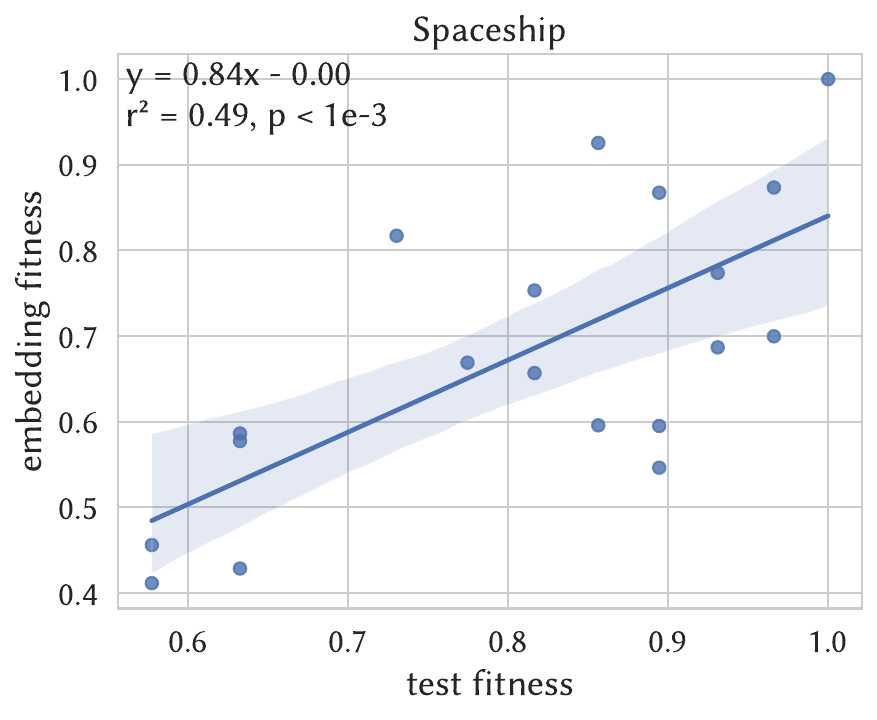}
    \caption{\qwenembedding with the individual sprites’ code as inputs followed by max-pooling aggregation.\label{fig:buggy-projects:qwen-sprite}}
  \end{subcaptionblock}
  \caption{\label{fig:buggy-projects}Linear least-squares regression between test and embedding space distances between student projects and example solution.}
\end{figure}

The previous research question demonstrated that the embeddings encode
relevant information about the programs. In this research question we
now evaluate whether this can be used to predict the correctness of a
program, even when the models are not fine-tuned for the specific task
due to the limited amount of available training data.
In our evaluation we observe a statistically significant linear
correlation~(Pearson’s \(r)\) between the test fitness~(\ie true
correctness as measured by the test suite) and the embedding fitness
for most base-project and embedding model combinations.
The slopes \(<1\) in most cases~(cf.\ \cref{fig:buggy-projects})
suggest that the embedding fitness tends to underestimate the true
functional correctness of the programs.

The test fitness correlates strongest with the embedding fitness of
the \ggnn model for the BoatRace project~(\(r^2 = \exnum{0.45}\),
\pValue[<]{e-3}, cf.~\cref{fig:buggy-projects:ggnn}). For the
Spaceship project on the other hand, the \qwen-per-sprite
embeddings perform best~(\(r^2 = \exnum{0.49}\), \pValue[<]{e-3},
cf.~\cref{fig:buggy-projects:qwen-sprite}). While the \ggnn model
performs best for FruitCatching and still exhibits the expected linear
correlation between embedding and test fitness~(\(p =
\exnum{0.018}\)), the linear relation between these two fitnesses is
weaker (\(r^2 = \exnum{0.13}\)).
The correlation for the FruitCatching project may be weaker since the
project allows for alternative solutions. Therefore, while some
student programs may be nearly correct~(\ie high test fitness), they
may still be structurally different to the example solution~(\ie low
embedding fitness).

For the BoatRace model, the \ggnn embeddings may outperform the two
\qwenembedding variants due to the sprite naming pre-training task.
Since the BoatRace project is based on a publicly available project
published by the CodeClubRik \Scratch
user~(cf.~\cref{sec:methods:project-remix-dataset}), our sprite naming
training dataset likely contains similar programs. Spaceship and
FruitCatching on the other hand are not based on publicly shared
base-projects. Therefore, our \ggnn model may be already somewhat
familiar with the structure of BoatRace-like programs whereas it is
unlikely to have seen the other two projects before.
Nevertheless, the embedding fitness of the \ggnn model still
correlates with statistical significance for the unseen
projects~(cf.~\cref{fig:buggy-projects:ggnn}), demonstrating that the
embeddings captured generalising semantic and structural information
about the programs during the pre-training on the sprite-naming task.

The difference in results between BoatRace and Spaceship may be
explained by their program structure. In principle, in both projects a
vehicle follows the mouse pointer through a labyrinth-like scene
towards a finish. However, for BoatRace the majority of the program’s
code is part of the \enquote{Boat} sprite. In Spaceship a similar
amount of code is spread out over the three sprites.
For the per-sprite embedding aggregation of BoatRace, the nearly empty
sprites might result in embeddings containing little information which
however overrides the information from the Boat sprite in the
max-pooling step. Therefore, \qwen-whole performs better than the
aggregated per-sprite embeddings~(whole program embedding: \(r^2 =
0.37\), per sprite: \(r^2 = 0.21\), cf.\
\cref{fig:buggy-projects:qwen-sprite,fig:buggy-projects:qwen-whole}).
Due to the better distribution of the code in the Spaceship project,
the individual sprite embeddings likely contain similar amounts of
information and thus retain more of it during the aggregation,
resulting in a similar performance for the two \qwenembedding
variants~(whole program: \(r^2 = 0.44\), per sprite: \(r^2 = 0.49\)).

\begin{summarybox}{4}
We observe a statistically significant correlation between true test
  fitness and embedding fitness for most project/embedding model
  combinations. This demonstrates that the embeddings capture semantic
  and structural information about the programs in the pre-training
  task. This information is applicable to other tasks without further
  fine-tuning.
\end{summarybox}

\subsubsection{Discussion}

The results for this research question demonstrate that the
pre-trained sprite naming embeddings can be applied to tasks unrelated
to the pre-training even in cases where task-specific fine-tuning is
not feasible, for example due to a limited dataset size. This suggests
that the development of future embedding-supported educational tools
could be eased by using pre-trained embeddings rather than having to
train task-specific models.

For example, the specific task of the embedding-based program
correctness estimation could also be integrated into part of automatic
program repair tools since the repeated test suite execution tends to
represent a considerable proportion of the overall repair
time~\cite{Schweikl2025RePurr}.
Similarly, combination of semantic and structural similarity between
projects could allow for the integration as part of a next-step hint
system guiding students towards an exercise solution. Current hint
systems tend to rely on only the structural differences between
student program and example solution~\cite{Fein2022Catnip}.

\subsection{RQ5: Program Embeddings as Student Progress Indicator}

\subsubsection{Results}

As demonstrated in RQ4, the embeddings can be used to estimate whether
a program is functionally similar to an example solution. While this
already could give some indication of the progress in an exercise, the
progress-variance-projection~\cite{Paassen2021Mapping}~(cf.~\cref{sec:setup:rq5-methods})
is specifically designed to map the embeddings into a two-dimensional
representation where the x-axis represents the progress between the
starting project at point \((0,0)\) and the example solution at
\((1,0)\).

A linear least-squares regression between the progress component of
the projection and the test fitness of the programs confirms that the
projection indeed shows the actual functional progress (\(p\)~values
for all models \(\ll 0.05\)).
The linear model fits worst for the \qwenembedding with max-pooled
sprite
embeddings~(\enquote{\qwen-per-sprite}, \mbox{\(r^2=\num{0.43607}\)}),
but considerably better for \qwenembedding with the full program as
input~(\enquote{\qwen-whole}, \mbox{\(r^2=\num{0.71373}\)}) and for
\ggnn~(\mbox{\(r^2=\num{0.73495}\)}).
Repeating the linear least-squares regression separately for the
programs of each student, we obtain a population of \(r^2\)
coefficients we can use for a statistical comparison between models. A
Friedman test~\cite{Friedman1937Use} shows that there is a significant
difference between populations~(\pValue{9.009376217182378e-09}). A
post-hoc Nemenyi test~\cite{Nemenyi1963Distribution} shows no
significant difference~(\mbox{\(\CD = \num{0.524}\)}) between \ggnn
and \qwen-whole. However, both models perform significantly better
than \qwen-per-sprite with large Vargha-Delaney effect
size~\cite{Vargha2000Critique}~(\ggnn compared to \qwen-per-sprite:
\(\hat{A} = \num{0.8125}\), \qwen-whole compared to \qwen-per-sprite:
\(\hat{A} = \num{0.809375}\)).

\begin{figure}[t]
  \centering \includegraphics[width=\linewidth]{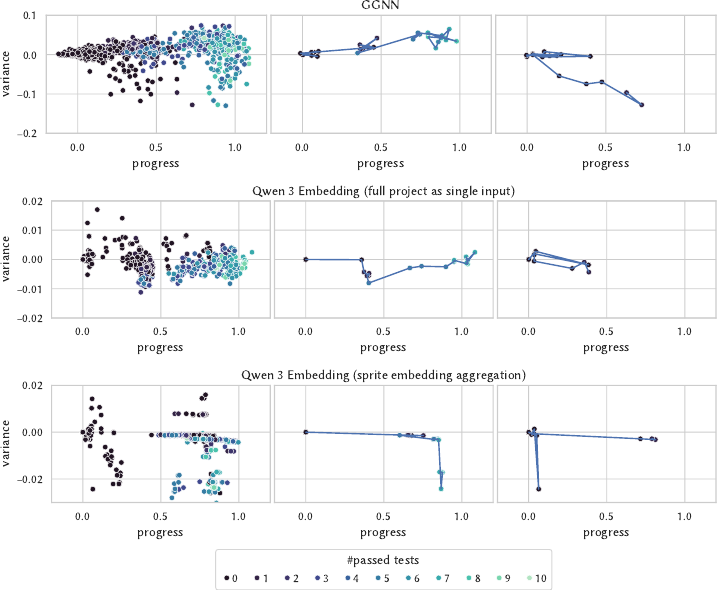}
  \caption{\label{fig:progress-variance:combined}Progress-variance-projection of student program snapshots for the whole class captured every minute.
    The example student shown in the middle followed a direct path towards the solution with some trial-and-error at the beginning, the middle, and towards the end.
    The student shown on the right got stuck and did not finish the tasks.
  }
\end{figure}

The correlation between test fitness and progress is highlighted in
\Cref{fig:progress-variance:combined}. The scatter plot of all student
programs shows that the number of passed tests generally increases as
the programs progress closer towards the solution. The large jump in
progression between projects close to the start to progress values
\(>0.5\) for \qwen-per-sprite might explain the lower \(r^2\)
coefficient for this model. This gap also makes the projection less
useful in practice, since the progressions for both shown example
students jump from the start state to a point already fairly close to
the solution. For the \qwen-whole variant and \ggnn, both students
progress through multiple intermediate steps.

The y-axis of the projection shows the variance and is intended to
highlight differences in programs with similar progression. Outliers
on the y-axis can indicate students no longer following the task. A
useful embedding projection should therefore map to a sufficient range
of variance values.
The projection based on the \ggnn model exhibits the highest
interquartile range~(\(\IQR=\num{0.02132402385944874}\)) for the
variance axis. For both \qwenembedding variants, the variance values
are an order of magnitude smaller~(whole program:
\(\IQR=\num{0.0032313232502529715}\), aggregated:
\(\IQR=\num{0.0029130429014551655}\)). In practice this is not
relevant however, since a visualisation similar to
\cref{fig:progress-variance:combined} can be scaled accordingly.
More relevant in practice is the overall spread of values on the
variance axis. For \ggnn the range is \(\num{9.559075315197754} \cdot
\IQR\) and \qwen-whole-program has a similar range of
\(\num{8.741470419450938} \cdot \IQR\).
When defining outliers as all points positioned \(1.5 \cdot \IQR\)
below/above the first/third quantile, the projection of \ggnn
embeddings results in \exnum{88} outliers, whereas for \qwen-whole
\exnum{50} outliers exist.

\begin{summarybox}{5}
Our results show that the pre-trained \ggnn sprite naming embeddings
  capture relevant transferrable program information that can be used
  to visualise the student progress without requiring further
  fine-tuning. In case no specifically trained \Scratch model is
  available, an embedding-\llm can be used instead with comparable
  performance.
\end{summarybox}

\subsubsection{Discussion}

In practice, not only the quality of the projection may be relevant
but also the computational effort required to generate it. Once the
\ggnn model has been pre-trained, embeddings can be computed locally
on a machine equipped only with a \cpu. Due to the large model size,
computing the \llm-based embeddings on the other hand is
computationally more expensive and requires a \gpu to support the
near-real-time computations required to enable in-classroom support
for teachers. While hosted \llm{}s accessible via a remote \api are
available, such \api{}s incur costs based on the usage and data
protection restrictions might apply.
Our results demonstrate that both approaches of using either a
dedicated small model or a pre-trained \llm can be used for the task
with similar results, making the approach more widely accessible by
allowing alternative implementations based on the individual compute
resource availability.

\section{Related Work}

The main related areas of research are either machine learning based
on visual aspects of code written in textual or visual programming
languages, or code embeddings specifically designed for visual
programming languages.

\subsection{Machine Learning Based on Visual Aspects of Code}

Most code embedding models designed for textual programming languages
receive the code in various textual~(\ie flat token sequence) or
structural formats~(\ie \AST, graphs). To understand whether such
established models can be applied in the \Scratch context, we
therefore ignored the visual aspects of the block-based programming
language in this paper.
Nevertheless, some embedding models for textual programming languages
include visual information, either as addition to the textual
representation~\cite{Mi2022Towards} or as
replacement~\cite{Chen2020Software,Kabore2023CodeGrid,Keller2021What}.
Using the notion of a grid-like layout of a source code file, the
bounding-boxes of potentially vulnerable code areas can be
determined~\cite{Li2022SySeVR}. The grid-like representation has also
been used to transform the source code into images by replacing the
characters with pixel colours based on the \initialism{ascii}
value~\cite{Chen2020Software}. To further abstract from the concrete
syntax, the inputs for
the \properNoun{CodeGrid}~\cite{Kabore2023CodeGrid} and
\properNoun{WySiSym}~\cite{Keller2021What} code clone and
vulnerability detection models are image with colours based on the
kind of the token~(\eg~identifier, keyword, literal). A similar
colouring approach has also been employed for automated code
readability analysis~\cite{Mi2022Towards}.

Even though the kinds of blocks in \Scratch already determine a
natural colouring of the pixels~(\eg green for expressions, blue for
blocks relating to the sprite movement), the grid-like image
generation approaches cannot be directly applied to \Scratch since its
workspace allows users to freely place blocks in two-dimensional space
independent of a grid. While screenshots of the code could be used
instead, additional challenges like image scaling and the possibility
of overlapping blocks need to be solved.
Due to the additional challenges, we did not investigate the impact of
including the visual aspects of \Scratch code for code embedding
models in this paper, but future research should explore this area.

\subsection{Code Embeddings for Visual Programming Languages}

Other initial work on code embeddings for \Scratch focusses on a
purely textual or structural representation of the code, even though
they do not yet use it to prompt \llm{}s. For example, the
\properNoun{ScratchRec} tool~\cite{Qi2022Scratch} uses a tokenised
representation of the program to create a program embedding which can
later be used to suggest similar projects to learners. A tokenisation
approach similar to our \neuralcodesum input
format~(cf.~\cref{sec:code-processing:tokens}) has been used to create
a code completion model for \Scratch~\cite{Griebl2023Applicability}.
One publication hints at using screenshots of \Scratch code to detect
bug patterns~\cite{Chai2023Towards}, but offers little detail about
processing steps or the used machine learning model.
All of these existing works use a specific code embedding model to
solve the task at hand, rather than comparing the performance of
different models.

The main focus of recent work on code embeddings for block-based
programming languages seems to be the conversion of the code into a
flat textual representation so that it can be fed into an \llm.
For example, the \properNoun{ChatScratch}
tool~\cite{Chen2024ChatScratch} allows students to get coding guidance
by automatically including the code in a syntax similar to
\scratchblocks in the prompts. The \llm’s output can be parsed back
into \Scratch blocks by heuristically matching similar block
types~\cite{Chen2024ChatScratch}. The \properNoun{Scratch
Copilot}~\cite{Druga2025Scratch} extends the \Scratch user interface
with a similar \llm-based coding help, but the publication does not
specify which code representation was used. A similar \llm-based
extension to the \Scratch user interface allows for the generation of
quizzes to encourage students to think about their
code~\cite{Druga2023Scratch}. While the latter two publications hint
at including the code in textual form into the prompt, they do not
specify the representation further.
Other works fully adopted the \scratchblocks format both as input and
expected output format to or from the
\llm~\cite{Fein2025LitterBox+,Fein2026Reasoning} to ask the \llm to
explain current issues in the code and suggest possible fixes.
Subsequently, we adopted the \scratchblocks format when prompting
\llm{}s in this paper~(cf.~\cref{sec:code-processing:llms}). The
\properNoun{Stitch} conversely uses the \Scratch-internal \json
representation as inputs for an \llm-based next-step hint generator,
but does not require the \llm to produce valid \json{}s as
output~\cite{Si2025Stitch}.

Since some \llm{}s are multi-modal, \ie they can accept text and
images as inputs, some recent work uses this feature to create
education-focussed tools for \Scratch.
For example, \properNoun{ScratchEval}~\cite{Fu2025ScratchEval}, uses
images of \Scratch code and a multiple-choice question that requires
reasoning about maths, logic, or graphics and use the \llm to
determine the solution. Similar to our RQ2, they evaluated the impact
of the natural language and found that images with descriptions in
Chinese resulted in roughly 2--3\,\% worse performance compared to
English~\cite{Fu2025ScratchEval}.
The \properNoun{VisionScratch} tool~\cite{Si2026VisionScratch} aims to
generate feedback for learners by feeding the code and a video
recording of the gameplay rather than the code into an \llm.

\Scratch seems to be the most common block-based programming language
which machine learning approaches are applies to. While the similar
\snap language is popular in educational contexts, no code embeddings
have been designed for it to the best of our knowledge.
Embedding models have also been applied to the \properNoun{Simulink}
language by transforming the graphs into flat textual representations
to generate the inputs for an
\llm~\cite{Luitel2024Requirements,Zhang2025Simulink}. Even though
\properNoun{Simulink} is a visual domain-specific language, it is
considerably different from \Scratch since the graph primarily
represents data flow between physical components and system states
rather than iterative statements.

\section{Conclusions}

In this paper we explored in an initial study how established code
embedding models can be applied to \Scratch. Our results for the
sprite-naming task~(RQs 1 and 2) clearly indicate that more structural
information about the program results in an improved classification
performance. However, when applying the models to tasks requiring
embeddings for the whole project~(RQs~3--5), a pre-trained
embedding-\llm performs similarly to our best-performing dedicated
\Scratch embedding model.

This indicates that the two kinds of models could be used in
complementary roles. \llm{}s might be suitable for higher-level tasks
where overall similarities between projects are relevant, like for
example the progress estimation~(RQ5) or estimating the difficulty of
exercises. As prior research shows, the \llm{}s’ large pre-training
datasets allow them to be used for tasks combining code context with
natural language explanations.
The \Scratch-specific models on the other hand might be more suitable
for programming assistance tools where structural insights are
required, for example next-step hint generators or bug-detectors.
Since they can be trained for specific tasks, the creation of
dedicated small models allow for less resource intensive deployments.
This might be relevant in both large- and small-scale deployments
where the required resources to host an \llm are not available or
externally hosted tools cannot be used due to the special care
required when working with children’s data.

Since this is the first study comparatively exploring code embeddings
for \Scratch, many opportunities for future improvements and
extensions arise.
The next steps to improve the models’ performance for the sprite
naming task could be an improved tokenisation approach. Similarly, the
current models could be extended with suitable changes to the input
format and model structure to work on whole programs rather than
individual sprites.
Based on related research for code embeddings on regular textual
languages and due to the multi-modality of \llm{}s, including more
\Scratch-specific information like for example block positions or
visual information seems like a promising direction for future
research.
More generally, when considering the event-based and concurrent
execution flows commonly found in \Scratch, an integration of these
concepts into the structure given to code embedding models might also
be of relevance beyond \Scratch when embedding the code of other
event-driven systems like games and game frameworks, mobile apps, or
distributed services.

\section*{Data Availability}

A replication package is provided online at \href{https://doi.org/10.5281/zenodo.21273160}{doi:10.5281/zenodo.21273159}.
It contains the sampled \Scratch projects used for model training and evaluation, our model implementations, the raw model inference results, and our evaluation scripts.

\printbibliography 

\end{document}